\documentclass{aa}  

\usepackage{pifont}
\usepackage{amsmath}
\usepackage{graphicx}
\usepackage{txfonts}
\usepackage{url}
\usepackage{subfigure}
\usepackage{longtable}
\usepackage{lscape}
\usepackage{natbib}

\newcommand{\HII}{H {\small{II}} }
\newcommand{\kms}{{\rm km~s}^{-1}}
\newcommand{\Msun} {\rm M_{\sun}}

\newcommand{\mjyb}{{\rm mJy~beam}^{-1}}

\newcommand{\dotsec}{\rlap.{''}}

\begin{document} 

   \title{N131: A dust bubble born from the disruption of a gas filament}
  % \title{The origin of the infrared dust bubble N131}
%   \subtitle{bubble N131}
    \authorrunning{C.-P. Zhang et al.}
    \titlerunning{The infrared dust bubble N131}

   \author{Chuan-Peng Zhang
          \inst{1,2,5,6}
          \and
          Guang-Xing Li\inst{2,3}
          \and 
          Friedrich Wyrowski\inst{2}
          \and
          Jun-Jie Wang\inst{1,5}
          \and
          Jing-Hua Yuan\inst{1}  
          \and
          Jin-Long Xu\inst{1,5}
          \and 
          Yan Gong\inst{2,4,6}
          \and
          Cosmos C. Yeh\inst{2}
          \and
          Karl M. Menten\inst{2}
          %\fnmsep
          %\thanks{Just to show the usage of the elements in the author field}
          }

   \institute{National Astronomical Observatories, Chinese Academy of Sciences, 100012 Beijing, PR China\\
   \email{cpzhang@nao.cas.cn}
   \and
    Max-Planck-Institut f\"ur Radioastronomie, Auf dem H\"ugel 69, 53121 Bonn, Germany    
     \and
    University Observatory Munich, Scheinerstrasse 1, D-81679 Munich, Germany 
    \and
    Purple Mountain Observatory \& Key Laboratory for Radio Astronomy, Chinese Academy of Sciences, 2 West Beijing Road, 210008, Nanjing, PR China
   \and
    NAOC-TU Joint Center for Astrophysics, 850000 Lhasa, PR China
    \and
    University of the Chinese Academy of Sciences, 100080 Beijing, PR China
            %\email{c.ptolemy@hipparch.uheaven.space}
             %\thanks{The university of heaven temporarily does not accept e-mails}
             }

   \date{Received XXX, XXX; accepted XXX, XXX}

% \abstract{}{}{}{}{} 
% 5 {} token are mandatory
 
  \abstract
  % context heading (optional)
  % {} leave it empty if necessary  
   {OB-type stars have strong ionizing radiation and drive energetic winds. The ultraviolet (UV) radiation from ionizing stars may heat dust and ionize gas to sweep up an expanding bubble shell. This shell may be the result of feedback leading to a new generation of stars.}
  % aims heading (mandatory)
   {N131 is an infrared dust bubble residing in a molecular filament. We study the formation and fragmentation of this bubble with multiwavelength dust and gas observations.}
  % methods heading (mandatory)
   {Towards the bubble N131, we analysed archival multiwavelength observations including 3.6, 4.5, 5.8, 8.0, 24, 70, 160, 250, 350, 500 $\mu$m, 1.1 mm, and 21 cm. In addition, we performed new observations of CO (2-1), CO (1-0), and $^{13}$CO (1-0) with the IRAM 30 m telescope.}
  % results heading (mandatory)
   {Multiwavelength dust and gas observations reveal a ring-like shell with compact fragments, two filamentary structures, and the secondary bubble N131-A. Bubble N131 is a rare object with a large hole at 24 $\mu$m and 21 cm in the direction of its centre.  The dust and gas clumps are compact and might have been compressed at the inner edge of the ring-like shell, while they are extended and might be pre-existing at the outer edge. The column density, excitation temperature, and velocity show a potentially hierarchical distribution from the inner to outer edge of the ring-like shell. We also detected the front and back sides of the secondary bubble N131-A  in the direction of its centre. The derived Lyman-continuum ionizing photon flux within N131-A is equivalent to an O9.5 star. Based on the above, we suggest that the bubble N131 might be triggered by the strong stellar winds from a group of massive stars inside the bubble.} 
  % conclusions heading (optional), leave it empty if necessary 
   {We propose a scenario in which the bubble N131 forms from the disruption of a gas filament by the expansion of the \HII region, strong stellar winds, and fragments under self-gravity. }


   \maketitle
%
%________________________________________________________________

\section{Introduction}    %% first-level sections will be auto-capitalized
\label{sect:intro}

Interstellar bubbles are very common objects, but their expansion is an important mechanism that  reshapes the morphology of the interstellar medium \citep{chur2006,chur2007,simp2012,Hou2014}. The formation of bubbles is still unclear \citep[e.g.][]{beau2010}. Generally, most of the bubbles exhibit an expanding \HII region enclosed within the bubble \citep{deha2010}. It is likely that bubbles are three-dimensional (3D) structures with a blueshifted front side and a redshifted back side, but the front and back sides are difficult to detect \citep[e.g.][]{Peng2010,s51,beau2010}. If a bubble is expanding, we should easily observe the expansion of its ring-like shell; however, just a few examples  show the signatures of expanding shells, i.e. N6 \citep{Yuan2014}, N68 \citep{n68}, and N131 \citep{n131}. Frequently, we observe pillars and shell-like structures from dust emission along the expansion direction \citep[e.g. the bubble CN88 in][]{chur2007}, which might provide clues to the origin of bubbles.  Ring-like shells may be triggered through the ``collect-and-collapse'' process \citep{whit1994}, the ``radiation driven implosion'' of pre-existing dense molecular clumps \citep{Lefloch1994,deha2010}, stellar winds \citep{Dyson1980}, and supernova explosions \citep{chur2006}, which take part in shaping the morphological structure of the bubble, such as pillars and shell-like structures. Young stellar objects (YSOs) and IRAS sources are often distributed among ring-like shells. These newly formed stars are likely triggered by the expansion of bubbles \citep{Castor1975,Weaver1977,wats2008,s51,Cappa2014}.

\begin{figure*}
\centering
\includegraphics[width=0.65\textwidth, angle=0]{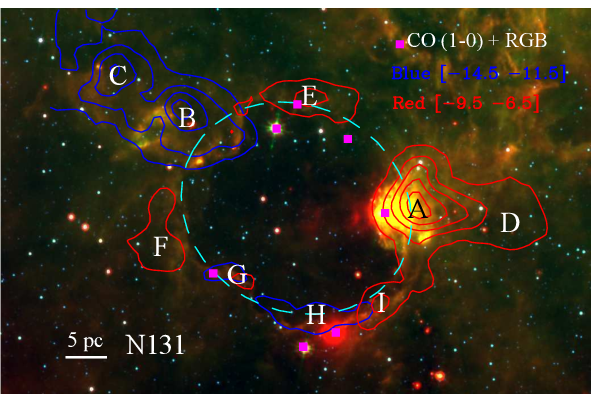}
\caption{Integrated intensity contours of the CO (1-0) line of the blueshifted and redshifted clouds superimposed on the RGB image with 24 $\mu$m (red), 8.0 $\mu$m (green), and 4.5 $\mu$m (blue). The CO (1-0) data is from \citet{n131}. The integration range is from -14.5 to -11.5 $\kms$ for the blueshifted cloud, and from -9.5 to -6.5 $\kms$ for the redshifted cloud. The symbols ``$\blacksquare$'', letters (A, B, ..., and I), and ellipse indicate the positions of eight IRAS point sources, nine molecular clumps, and ring-like shell of the bubble, respectively. }
\label{Fig:rgb}
\end{figure*}

\begin{figure*}
\centering
\includegraphics[width=0.99\textwidth, angle=0]{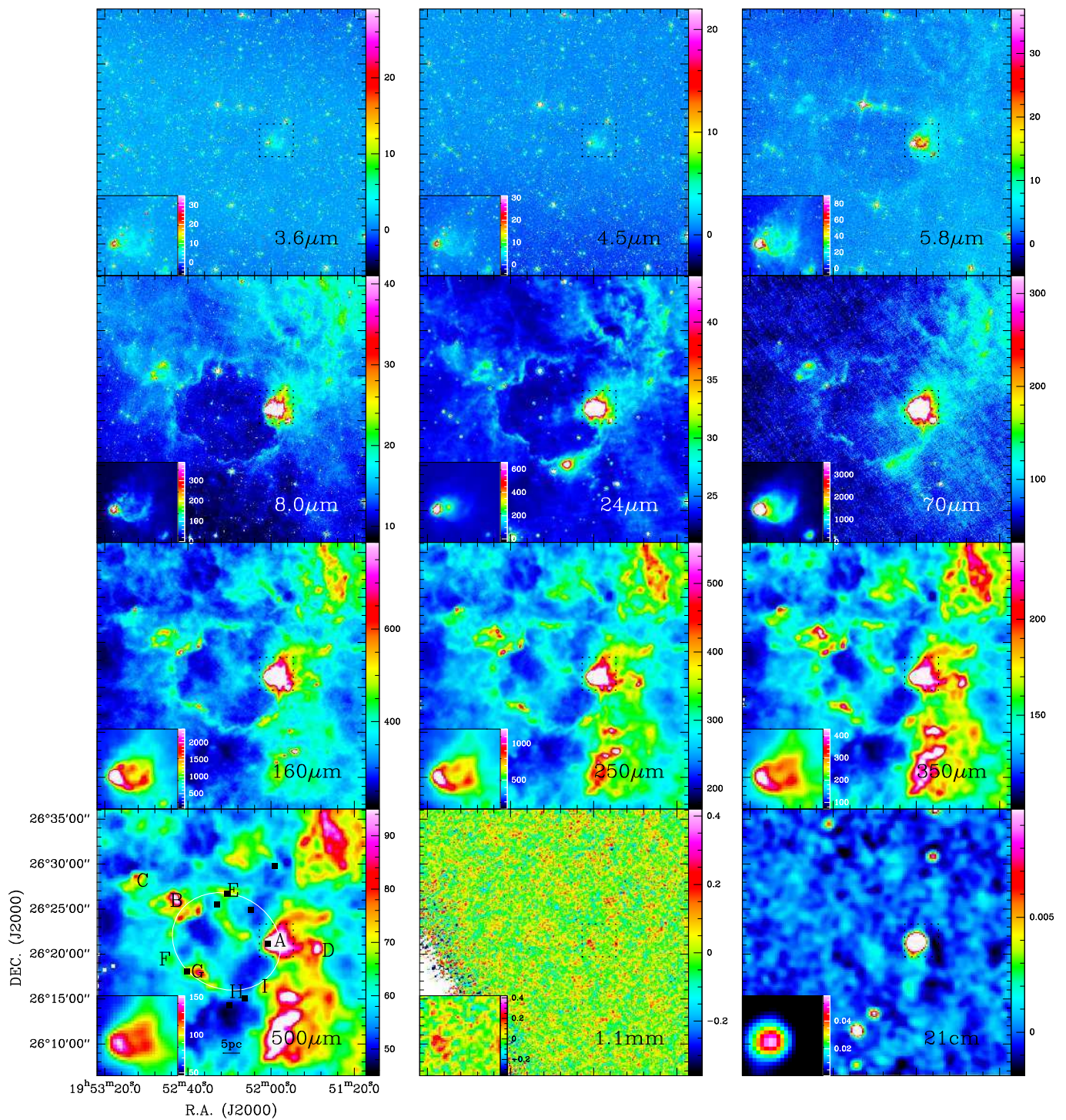}
\caption{Multiwavelength images of the bubble N131 from 3.6 $\mu$m to 21 cm. The saturated region within the dashed square is shown in the bottom-left corner of each subfigure, as a zoom in with adjusted intensity scale. The unit of each colour bar is, respectively, MJy sr$^{-1}$ for 3.6, 4.5, 5.8, 8.0, 24, 70, 160, 250, 350, 500 $\mu$m, Jy beam$^{-1}$ for 1.1 mm, and 21 cm. In the 500 $\mu$m subfigure, the symbols ``$\blacksquare$'', letters (A, B, ..., and I), and ellipse indicate the positions of eight IRAS point sources, nine molecular clumps, and the ring-like shell of the bubble, respectively. }
\label{Fig:irbubble}
\end{figure*}

\begin{figure*}
\centering
\includegraphics[width=0.60\textwidth, angle=0]{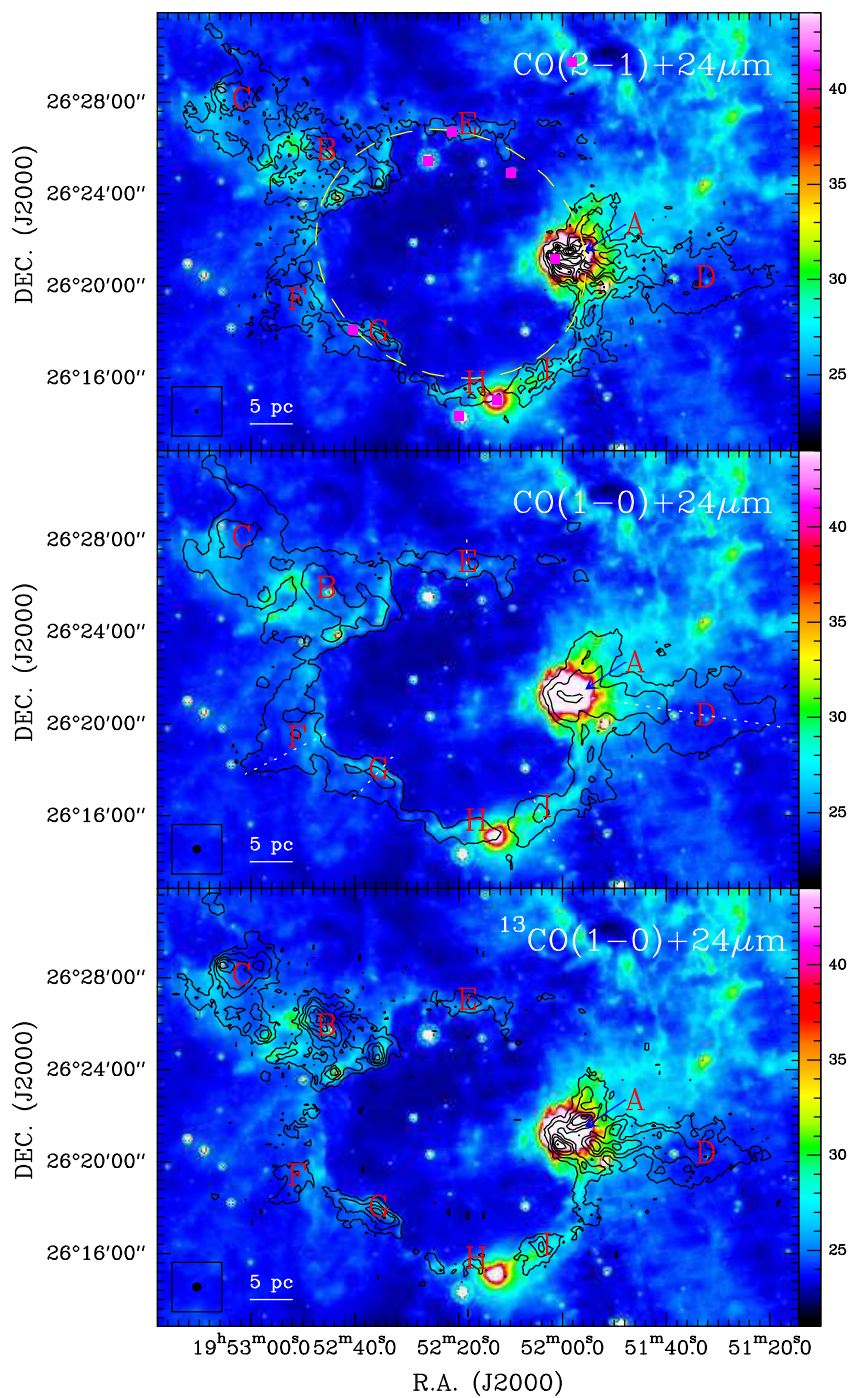}
\caption{Integrated intensity maps of the CO (2-1) ($upper$), CO (1-0) ($middle$), and $^{13}$CO (1-0) ($lower$) lines with velocity range from -14.5 to -6.5 $\kms$ superimposed on 24 $\mu$m emission. The contour levels start are at 3$\sigma$ in steps of 5$\sigma$ for CO (2-1) ($\sigma = 1.7\, {\rm K}(T^*_{\rm A})\, \kms$), CO (1-0) ($\sigma = 1.5\, {\rm K}(T^*_{\rm A})\, \kms$), and $^{13}$CO (1-0) ($\sigma = 0.4\, {\rm K}(T^*_{\rm A})\, \kms$). The beam size of each subfigure is indicated at the bottom-left corner. The symbols ``$\blacksquare$'', letters (A, B, ..., and I), and ellipse indicate the positions of  eight IRAS point sources, nine molecular clumps, and ring-like shell of the bubble, respectively. The unit of each colour bar is in MJy sr$^{-1}$. The dotted lines show the cutting direction of the slice profiles in Fig. \ref{Fig:slice} and the PV diagram in Fig. \ref{Fig:pv}.}
\label{Fig:int-map}
\end{figure*}

\begin{table*}
\caption{Parameters of fragments determined from \textit{Gaussclumps} analysis}.
\label{tab_parameter} \centering \scriptsize
\begin{tabular}{ccccccccccccc}
\hline \hline
No.     &       Offsets         &       Gaussian Size   &
\multicolumn{2}{c}{FWHM} & \multicolumn{2}{c}{Deconv. FWHM}     &       $N_{\rm H_2}$
&       $M$     &    $V_{\rm lsr}$    &    $\Delta V$   &    $M_{\rm vir}$   &  $\alpha_{\rm vir}$ \\
 &   $'',\,''$   &   $''\,\times\,''$   &   $''$   &   pc   &   $''$   &   pc   &  
$\rm 10^{21}cm^{-2}$   &   $\Msun$ & $\kms$ & $\kms$  & $\Msun$  \\
(1) & (2) & (3) & (4) & (5) & (6) & (7) & (8) & (9) & (10) &  (11) & (12) & (13) \\
\hline
1       &       (68.8,\,-95.9)  &       30.8    $\,\times\,$    61.2    &       38.9    &       1.6     &       31.7    &       1.3     &       21.2    &       894     &       -9.98(0.01)     &       1.60(0.02)      &       521     &       0.58    \\
2       &       (13.0,\,-31.5)  &       65.5    $\,\times\,$    42.5    &       50.4    &       2.1     &       45.1    &       1.9     &       16.3    &       1160    &       -9.10(0.02)     &       2.30(0.03)      &       1401    &       1.21    \\
3       &       (541.2,\,128.8) &       53.3    $\,\times\,$    37.1    &       43.0    &       1.8     &       36.7    &       1.5     &       15.5    &       800     &       -11.50(0.01)    &       1.68(0.03)      &       636     &       0.80    \\
4       &       (685.1,\,224.0) &       71.3    $\,\times\,$    132.1   &       88.7    &       3.7     &       85.8    &       3.6     &       14.9    &       3267    &       -12.28(0.01)    &       1.90(0.03)      &       1684    &       0.52    \\
5       &       (948.3,\,368.2) &       46.2    $\,\times\,$    71.1    &       54.8    &       2.3     &       49.9    &       2.1     &       14.5    &       1217    &       -12.87(0.01)    &       1.68(0.03)      &       811     &       0.67    \\
6       &       (-11.4,\,-88.1) &       82.8    $\,\times\,$    37.4    &       48.2    &       2.0     &       42.6    &       1.8     &       13.4    &       872     &       -8.98(0.01)     &       1.59(0.03)      &       637     &       0.73    \\
7       &       (660.9,\,88.5)  &       33.6    $\,\times\,$    70.1    &       42.9    &       1.8     &       36.5    &       1.5     &       13.3    &       682     &       -11.42(0.01)    &       1.39(0.03)      &       436     &       0.64    \\
8       &       (837.0,\,192.5) &       36.7    $\,\times\,$    41.4    &       38.8    &       1.6     &       31.6    &       1.3     &       12.3    &       519     &       -12.28(0.01)    &       1.01(0.02)      &       207     &       0.40    \\
9       &       (-75.3,\,-103.8)        &       28.8    $\,\times\,$    68.2    &       40.0    &       1.7     &       33.0    &       1.4     &       12.0    &       538     &       -9.21(0.02)     &       2.23(0.04)      &       1042    &       1.94    \\
10      &       (549.0,\,-271.9)        &       37.3    $\,\times\,$    116.6   &       50.2    &       2.1     &       44.9    &       1.9     &       11.1    &       779     &       -10.73(0.01)    &       1.35(0.02)      &       478     &       0.61    \\
11      &       (-67.2,\,-55.6) &       36.8    $\,\times\,$    25.6    &       29.7    &       1.2     &       19.4    &       0.8     &       10.1    &       251     &       -8.88(0.02)     &       1.98(0.04)      &       613     &       2.45    \\
12      &       (117.0,\,-367.6)        &       54.1    $\,\times\,$    53.2    &       53.6    &       2.2     &       48.7    &       2.0     &       8.1     &       648     &       -7.70(0.02)     &       1.78(0.05)      &       890     &       1.37    \\
13      &       (845.3,\,360.4) &       72.7    $\,\times\,$    107.7   &       88.5    &       3.7     &       85.6    &       3.6     &       8.0     &       1747    &       -11.74(0.02)    &       1.69(0.04)      &       1324    &       0.76    \\
14      &       (4.9,\,-151.8)  &       44.5    $\,\times\,$    22.7    &       28.6    &       1.2     &       17.7    &       0.7     &       7.5     &       171     &       -9.60(0.01)     &       1.24(0.03)      &       230     &       1.34    \\
15      &       (4.9,\,32.4)    &       39.1    $\,\times\,$    65.6    &       47.5    &       2.0     &       41.8    &       1.7     &       7.2     &       455     &       -8.81(0.01)     &       1.23(0.03)      &       380     &       0.84    \\
16      &       (285.0,\,272.3) &       163.4   $\,\times\,$    45.0    &       61.3    &       2.6     &       57.0    &       2.4     &       7.0     &       737     &       -8.51(0.04)     &       1.72(0.08)      &       958     &       1.30    \\
17      &       (-203.3,\,-119.7)       &       86.7    $\,\times\,$    96.4    &       91.4    &       3.8     &       88.6    &       3.7     &       6.9     &       1618    &       -10.42(0.02)    &       1.71(0.05)      &       1409    &       0.87    \\
18      &       (541.0,\,168.4) &       38.4    $\,\times\,$    104.7   &       52.2    &       2.2     &       47.1    &       2.0     &       6.5     &       497     &       -11.02(0.03)    &       1.95(0.06)      &       1046    &       2.11    \\
19      &       (780.9,\,136.3) &       94.3    $\,\times\,$    45.0    &       57.4    &       2.4     &       52.8    &       2.2     &       6.3     &       577     &       -11.26(0.01)    &       0.88(0.04)      &       232     &       0.40    \\
20      &       (-35.0,\,-159.7)        &       22.5    $\,\times\,$    43.2    &       28.2    &       1.2     &       17.0    &       0.7     &       6.2     &       137     &       -9.27(0.01)     &       1.10(0.03)      &       179     &       1.30    \\
21      &       (612.0,\,124.4) &       53.5    $\,\times\,$    79.4    &       30.1    &       1.3     &       19.9    &       0.8     &       5.8     &       146     &       -10.99(0.02)    &       1.26(0.04)      &       252     &       1.73    \\
22      &       (861.0,\,248.3) &       69.1    $\,\times\,$    46.2    &       54.3    &       2.3     &       49.4    &       2.1     &       5.1     &       423     &       -11.62(0.02)    &       1.40(0.04)      &       561     &       1.33    \\
23      &       (-99.1,\,-39.6) &       37.9    $\,\times\,$    28.7    &       32.4    &       1.4     &       23.3    &       1.0     &       5.0     &       145     &       -8.43(0.03)     &       1.85(0.09)      &       580     &       3.98    \\
24      &       (-307.1,\,-127.6)       &       42.6    $\,\times\,$    150.4   &       58.0    &       2.4     &       53.4    &       2.2     &       4.9     &       461     &       -10.04(0.02)    &       1.66(0.04)      &       843     &       1.83    \\
25      &       (-11.0,\,-279.6)        &       98.6    $\,\times\,$    64.4    &       79.7    &       3.3     &       76.5    &       3.2     &       4.8     &       858     &       -10.38(0.05)    &       1.95(0.11)      &       1591    &       1.85    \\
26      &       (749.0,\,-175.7)        &       49.3    $\,\times\,$    70.9    &       57.1    &       2.4     &       52.5    &       2.2     &       4.6     &       423     &       -9.89(0.03)     &       1.51(0.07)      &       688     &       1.62    \\
27      &       (1013.0,\,288.3)        &       65.7    $\,\times\,$    69.9    &       65.8    &       2.7     &       61.8    &       2.6     &       4.4     &       530     &       -12.95(0.02)    &       1.33(0.06)      &       615     &       1.16    \\
28      &       (-371.1,\,-127.7)       &       43.7    $\,\times\,$    92.2    &       60.2    &       2.5     &       55.8    &       2.3     &       4.3     &       433     &       -9.96(0.04)     &       1.64(0.08)      &       854     &       1.97    \\
29      &       (924.9,\,312.4) &       25.1    $\,\times\,$    96.0    &       35.2    &       1.5     &       27.0    &       1.1     &       4.2     &       145     &       -11.89(0.02)    &       1.43(0.04)      &       379     &       2.62    \\
30      &       (588.9,\,208.4) &       39.6    $\,\times\,$    47.9    &       43.1    &       1.8     &       36.8    &       1.5     &       4.1     &       212     &       -11.90(0.04)    &       1.45(0.10)      &       476     &       2.24    \\
31      &       (636.9,\,272.4) &       44.0    $\,\times\,$    23.7    &       29.5    &       1.2     &       19.1    &       0.8     &       3.8     &       92      &       -12.44(0.02)    &       0.99(0.04)      &       151     &       1.64    \\\hline
\end{tabular}
\tablefoot{Column (1) lists the Gaussian fragment number. Column (2) lists the position offsets from absolute coordinate $\alpha(J2000) = 19^{\rm h}51^{\rm m}55\rlap.{\rm ^s}212$ and $\delta(J2000 )= 26^{\circ}22'21\rlap.{''}04$. Column (3) lists the Gaussian size derived from $Gaussclumps$. Columns (4) -- (7) list the FWHM and deconvolved FWHM. Column (8) lists the extracted peak column density using $Gaussclumps$. Column (9) lists the obtained fragment masses. Columns (10) -- (11) list the velocities and line widths from Gaussian fit of $^{13}$CO (1-0). Column (12) lists the virial mass. And Column (13) lists the virial parameters $\alpha_{\rm vir} = M_{\rm vir}/M$.}
\end{table*}

Is there any connection between the formations of bubbles and filaments? Filamentary structures in molecular clouds are ubiquitous in the Milky Way, and have been identified from subparsec to kpc scale \citep{Wang2011,Jackson2010,Li2013,Peretto2014,Ragan2014,Goodman2014}. On the molecular cloud scale, filaments may form through converging flows \citep{Csengeri2011}, shear \citep{Smith2014}, the collision of shocked sheets \citep{Padoan2001}, instabilities in self-gravitating sheets \citep{Nagai1998}, or turbulence. They can evolve to form stars \citep{Kainulainen2013,Gomez2011,Wang2014}, and in the final stage the star formation process should back-react on the filamentary structures. This feedback can either trigger the formation of the next generation of stars \citep{Whitworth1994} or disrupt the whole structure. However this feedback has not been studied in detail. If a cluster of massive stars resides within a filament, the strong stellar winds from these stars may disrupt the filament's structure. Such a scenario might be ongoing in the dust bubble N131 and its associated molecular clouds.

A promising object that might provide new insight into the formation of bubbles and their relation to filaments is the infrared dust bubble N131. Based on previous work from \citet{n131}, we present its composite colour image in Fig. \ref{Fig:rgb}. This bubble was selected from the catalogue  published by \citet{chur2006}. Its intriguing molecular structure was first observed and investigated with three $\rm CO\,(1-0)$ isotopic variants by \citet{n131}. Bubble N131  has an inner minor radius of 13.0 pc and inner major radius of 15.0 pc at a kinematic distance of 8.6 kpc with a centre coordinate of R.A.(J2000) = 19$\rm ^h$52$\rm ^m$21$\rm ^s$.5, DEC.(J2000) = +26$^{\circ}$21$'$24$''$.0 \citep{n131}. The CO emission is well correlated with a ring-like shell of Spitzer 24 and 8.0 $\mu$m emission. There are two giant elongated molecular clumps (named clumps AD and BC) appearing on opposite sides of the ring-like shell of N131 (see Fig. \ref{Fig:rgb}). Clump BC at the north-east location has a velocity range of [-14.5 -11.5] $\kms$, while  clump AD at the south-west location is in the range of [-9.5 -6.5] $\kms$. In addition, there is a huge cavity inside the bubble seen in the 24 $\mu$m and 1.4 GHz emission, which suggests that the ionized gas and hot dust may be evacuated by strong stellar winds. There are seven IRAS point sources located along the ring-like shell and well correlated with several molecular clumps. By revealing the distributions of YSOs, \citet{n131} found that there are 15 ionizing stars located within the bubble, and 63 YSOs surrounding the bubble. These may be the result of an expansion of strong stellar winds within the bubble.

The dynamics of bubble N131 is very intriguing, especially the two giant molecular filaments with blueshifted and redshifted  velocity components \citep{n131}. However, the origin of its dynamical properties is not known. Investigating physical parameters such as the density, temperature, and velocity will help us to understand the formation of bubbles because these parameters may show a hierarchical condition of the associated clumps. By studying the morphology we might be able to understand its past. Bubble N131 is an excellent object for carrying out such a project. In this work, we investigate the formation and evolution of N131, using higher angular resolution CO observations, along with archival 3.6 $\mu$m to 21 cm  continuum data. In Section \ref{sect:obser} we describe the observations and data reduction. In Section \ref{sect:analysis} we analyse the observational results from different wavelengths for the dust and gas clumps. In Section \ref{sec:discussion} we study the fragmentation and subsequent star formation of the bubble N131. In Section \ref{sect:summary} the results are summarized.

 %\normalsize
 %\large

\section{Observations and data}
\label{sect:obser}

\subsection{CO observations}

We simultaneously carried out CO (2-1), CO (1-0), and $^{13}$CO (1-0) observations during 2014 April 17-20 using the IRAM 30 m telescope\footnote{Based on observations carried out with the IRAM 30m Telescope. IRAM is supported by INSU/CNRS (France), MPG (Germany) and IGN (Spain).} on Pico Veleta, Spain. The E90 and E230 bands of the new Eight MIxer Receiver (EMIR) covered the three lines. The receiver was tuned to cover the frequency ranges 110.186 to 112.006 GHz for $^{13}$CO (1-0), 113.466 to 115.286 GHz for CO (1-0), and 229.678 to 231.498 GHz for CO (2-1) observations. The Fast Fourier Transform Spectrometer (FFTS) backends were  set at 50 kHz resolution.

For  the CO (2-1), CO (1-0), and $^{13}$CO (1-0) lines, the half-power beam width (HPBW) was 11.3$''$, 22.5$''$, and 23.5$''$, the main beam efficiency ($B_{\rm eff}$) was 59\%, 78\%, and 78\%, and the forward efficiency ($F_{\rm eff}$) of the IRAM 30 m telescope was 92\%, 94\%, and 94\%, respectively. The relation between main beam temperature ($T_{\rm mb}$) and antenna temperature ($T^*_{\rm A}$) is $T_{\rm mb} = (F_{\rm eff} / B_{\rm eff}) \times T^*_{\rm A}$. The on-the-fly mapping mode was used to scan the bubble shell in two orthogonal directions, to reduce striping on the maps. Given limited observing time, the sampling step was set as 9.3$''$, which meets the Nyquist sampling theorem for CO (1-0) and $^{13}$CO (1-0), yet not for CO (2-1). However, this was ameliorated by scanning in the two orthogonal directions.

Calibration scans, pointing, and focus were done on a regular basis to assure correct calibration. Calibration scans were done at the beginning of each subscan. A pointing was done about every hour, and a focus scan every three hours with more scans performed around sunset and sunrise. The flux calibration is expected to be accurate within 10\%. The GILDAS\footnote{http://www.iram.fr/IRAMFR/GILDAS/} and MIRIAD\footnote{http://www.cfa.harvard.edu/sma/miriad/} software packages were used to reduce the observational data.

\subsection{Archival data}

In Fig. \ref{Fig:irbubble}, multiwavelength observations are shown for bubble N131 from 3.6 $\mu$m to 21 cm. The combined data comprise GLIMPSE 3.6, 4.5, 5.8, 8.0 $\mu$m \citep{benj2003,chur2009}; MIPSGAL 24 $\mu$m \citep{care2009}; PACS 70 and 160 $\mu$m \citep{Poglitsch2010}; SPIRE 250, 350, and 500 $\mu$m \citep{Griffin2010}; BGPS 1.1 mm \citep{Aguirre2011,Ginsburg2013}; and NVSS 21 cm \citep{cond1998}. These data were direct archive products. The brightest region (or saturated region) with dashed square was zoomed in, and then shown in the bottom left corner of each subfigure as a zoom-in with adjusted intensity scale.

\section{Analysis and results}
\label{sect:analysis}
\subsection{Multiwavelength dust observations}

The 3.6 and 4.5 $\mu$m emission mostly originates from some stars associated with  N131 or bright field stars. The emission at both wavelengths does not delineate the shell structure of the bubble, except for several bright stars in Fig. \ref{Fig:irbubble}. These bright stars are associated with IRAS sources \citep{n131}. Zoomed in the dashed square, the 3.6 and 4.5 $\mu$m emission shows a cometary morphology (named N131-A) with a wide angle tail. The 5.8 and 8.0 $\mu$m emission originates mainly from polycyclic aromatic hydrocarbons (PAHs) \citep{wats2008}, which are excited by the photodissociation region (PDR) at the interface between the wind-blown bubble and the ambient interstellar medium. The shell structure begins to be visible at 5.8 $\mu$m, and becomes prominent at 8.0 $\mu$m.

The 24 and 70 $\mu$m emission is mostly produced by relatively hot dust. The 70 $\mu$m emission corresponds to hot dust emission of $\geqslant$ 40 K \citep{Faimali2012}, and partly to cool dust emission \citep{Anderson2012}. \citet{deha2010} found that about 98\% of the bubbles exhibit extended and prominent 24 $\mu$m emission enclosed within the bubble. In Fig. \ref{Fig:irbubble}, however, the 24 $\mu$m emission clearly delineates the ring-like shell of bubble N131. Inside the bubble only a weak patch of 24 $\mu$m emission is found. It is associated with an unrelated velocity component at 25.0 $\kms$ \citep{n131}. So, the bubble shows a large cavity at 24 $\mu$m inside the bubble. The inner edge of the shell is brighter than the outer edge at 24 $\mu$m, which shows an apparent intensity gradient (see Section \ref{sect:flux_var}). Along the ring-like shell at 24 $\mu$m, there are also several bright point sources associated with IRAS sources \citep{n131}. This is likely from feedback of the shocked stellar winds within the bubble. For N131-A region (see Fig. \ref{Fig:irbubble}) rescaled within the dashed square, the morphology of the 24 $\mu$m, showing a bright and compact spot, is not similar to 3.6 - 8.0 $\mu$m emission. We suggest that N131-A is actually a secondary bubble (see Section \ref{sect:small_bubble}). At 70 $\mu$m, the emission from the ring-like shell becomes weak, yet still visible. N131-A in the zoomed square becomes very bright with extended tails at 70 $\mu$m.

The 160, 250, 350, and 500 $\mu$m emission originates from cool dust with an average temperature of 26 K along the PDR \citep{Anderson2012}. This cool dust may be from the extended envelope of the ring-like shell. From 160 to 500 $\mu$m, the dust structure becomes more extended, partly because of the  lower spatial resolution and because  the longer wavelengths trace colder and more diffuse regions than the PDR shell. Next to the west side of the bubble there is a large ridge extending from north to south. This ridge has no detected velocity component associated with the bubble, so it may be partly from an unrelated background or foreground. Inside the bubble, one dust clump is visible. This clump is associated with another CO velocity component  at 25.0 $\kms$ \citep{n131}, excluding the possibility of it being the front side and back side of a 3D bubble structure \citep[e.g. ][]{s51}.  Within the zoomed square,  N131-A has two visible tails. Outside the square, N131-A also has two filamentary clumps towards the west. We propose that it is due to feedback of the stellar winds inside the bubble.

The 1.1 mm emission is dominated by cold dust. At limited sensitivity, we can just see an extended clump associated with N131-A in Fig. \ref{Fig:irbubble}. The 21 cm emission is mainly from free-free emission, which traces \HII regions. The 21 cm emission is also coincident with N131-A within the square. The peak of the 21 cm emission is located at the peak of the secondary bright point source (see Section \ref{sect:small_bubble}).

\subsection{Fragment extraction}
\label{sect:fragment_extraction}

To study the fragmentation in the ring-like shell of the bubble N131, the \textit{Gaussclumps} algorithm in GILDAS was used to extract dense fragments \citep{Stutzki1990,Kramer1998}. \textit{Gaussclumps} identifies local maxima in images and fits them with Gaussian intensity distributions. When fitting a Gaussian to the maximum, a modified chi-squared function is minimized. Three ``stiffness'' parameters control the fitting, ensuring that a local clump is fitted and subtracted. We only considered those fragments with the peaks of column density above 5$\sigma$ and with the fitted Gaussian FWHMs larger than their beam size. The derived parameters are listed in Table \ref{tab_parameter}, namely position, size, column density, and mass. The column densities and masses are calculated in Section \ref{sect:column_desity}.

\subsection{Morphology of molecular clumps}

Earlier, lower-resolution $^{12}$CO (1-0) observations have identified nine extended molecular clumps surrounding the shell of N131 \citep[labelled  A - I; see Fig. 2 of][]{n131}. The nine clumps were found surrounding a cavity which opens towards the north. An apparent intensity gradient from the inner to the outer edge of the ring-like shell was revealed in addition to two giant elongated molecular clumps at the opposite sides of the ring-like shell of N131. For the two molecular clumps, the brightest positions in $^{12}$CO (1-0) distribution are located at the ring-like shell, and then the emission gradually becomes diffuse outwards. Owing to density differences in pre-existing gas, it is likely that the strong expansion just moved into low densities, but expansion was slowed down at high densities.

In Fig. \ref{Fig:int-map}, we present three higher angular resolution integrated intensity maps of CO (2-1), CO (1-0), and $^{13}$CO (1-0). At large scales, the higher resolution observations have a similar morphology to previous molecular observations. At small scales, there are some intriguing differences between them. Clump A is resolved into several small fragments, which show filamentary structures elongating along the molecular clump (see also  Fig. \ref{Fig:3color} zoomed in). Among the locations of the small fragments within clump A, there is an elongated cavity in the east-west direction. Clump B also breaks up into several small fragments along the northeast-southwest direction.

\subsection{Comparison between dust and gas clumps}
\label{sect:differences}
The molecular CO emission in Fig. \ref{Fig:int-map} is correlated with the cold dust emission distribution in Fig. \ref{Fig:irbubble} \citep[see also Fig. 2 in][]{n131}. Between clumps A and E, the ring-like shells open towards the north-west direction where no dust and gas emission has been detected, while two IRAS point sources are detected nearby. Towards clumps C, D, and F, the dust emission is relatively weak; however, their CO emission is stronger. We suggest that the clumps C, D, and F resemble infrared dark clouds (IRDCs) 
\citep[e.g.][]{Rathborne2006,Simon2006}. The three clumps are relatively far away from the bubble centre. Furthermore, the ring-like shells of the CO emission have a bright inner edge and a dark outer edge of dust emission.

The IRAS point source 19499+2613 is close to the eastern part of  the secondary bubble N131-A, and is located at the head position of cometary clump A. As seen in Fig. \ref{Fig:int-map}, like the dust emission, the CO emission is extended to the west, although on a much larger scale. The cometary tails are the shell of the secondary bubble N131-A (see Fig. \ref{Fig:3color}). However, the material in this shell might have been partly influenced by the strong stellar winds of N131.

\subsection{Flux distribution on the ring-like shell}
\label{sect:flux_var}

Figure \ref{Fig:int-map} shows a good correlation between the 24 $\mu$m and CO (1-0) morphology. Slice profiles of the 24 $\mu$m and the CO (1-0) emission are shown in Fig. \ref{Fig:slice} through the clumps AD, E, F, and G. We found a steep rise at the inner edge of the bubble and a gradual fall at the outer edge of the bubble. This steep rise indicates that the inner part of the bubble has been compressed by the expansion from the stellar wind. Bubble N131 is very hollow, and most of the material within the bubble has been likely transported onto the ring-like shell. The gradual fall indicates that outside the bubble the molecular gas is pre-existing and that there has been little interaction with the stellar winds within the bubble.

\subsection{Position-velocity diagram}
\label{sect:3D}

\begin{figure*}
\centering
\includegraphics[width=0.45\textwidth, angle=0]{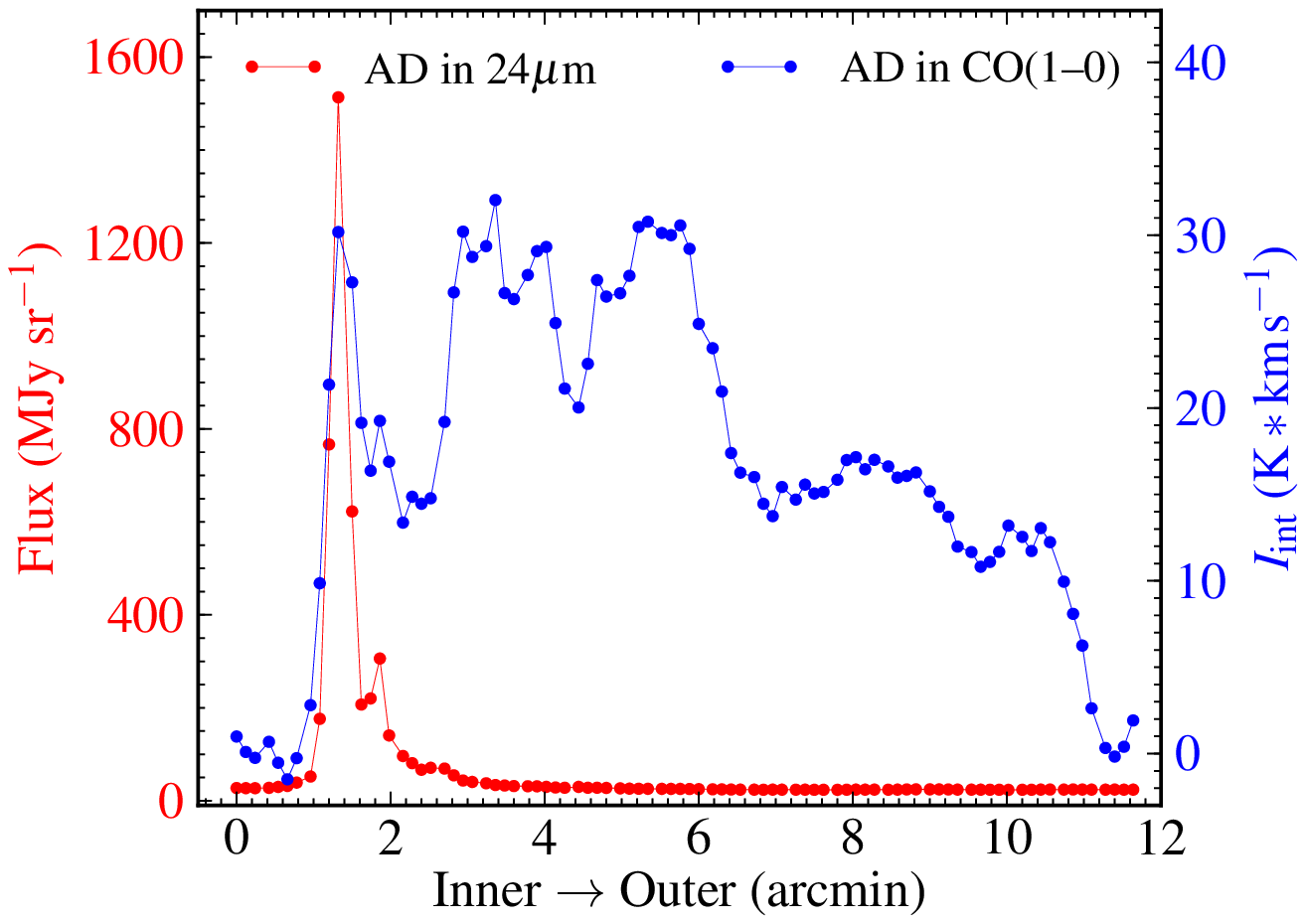}
\includegraphics[width=0.45\textwidth, angle=0]{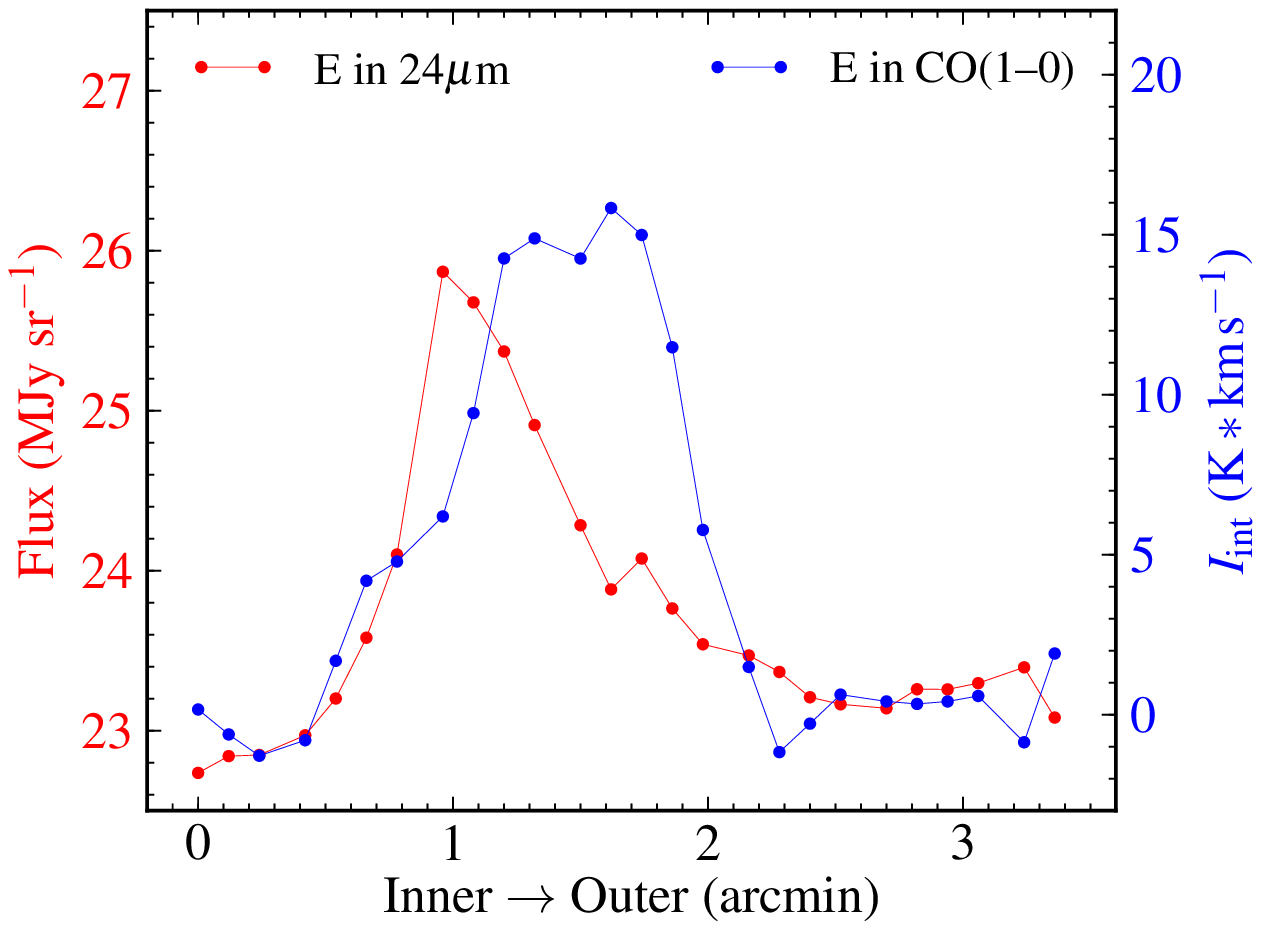}
\includegraphics[width=0.45\textwidth, angle=0]{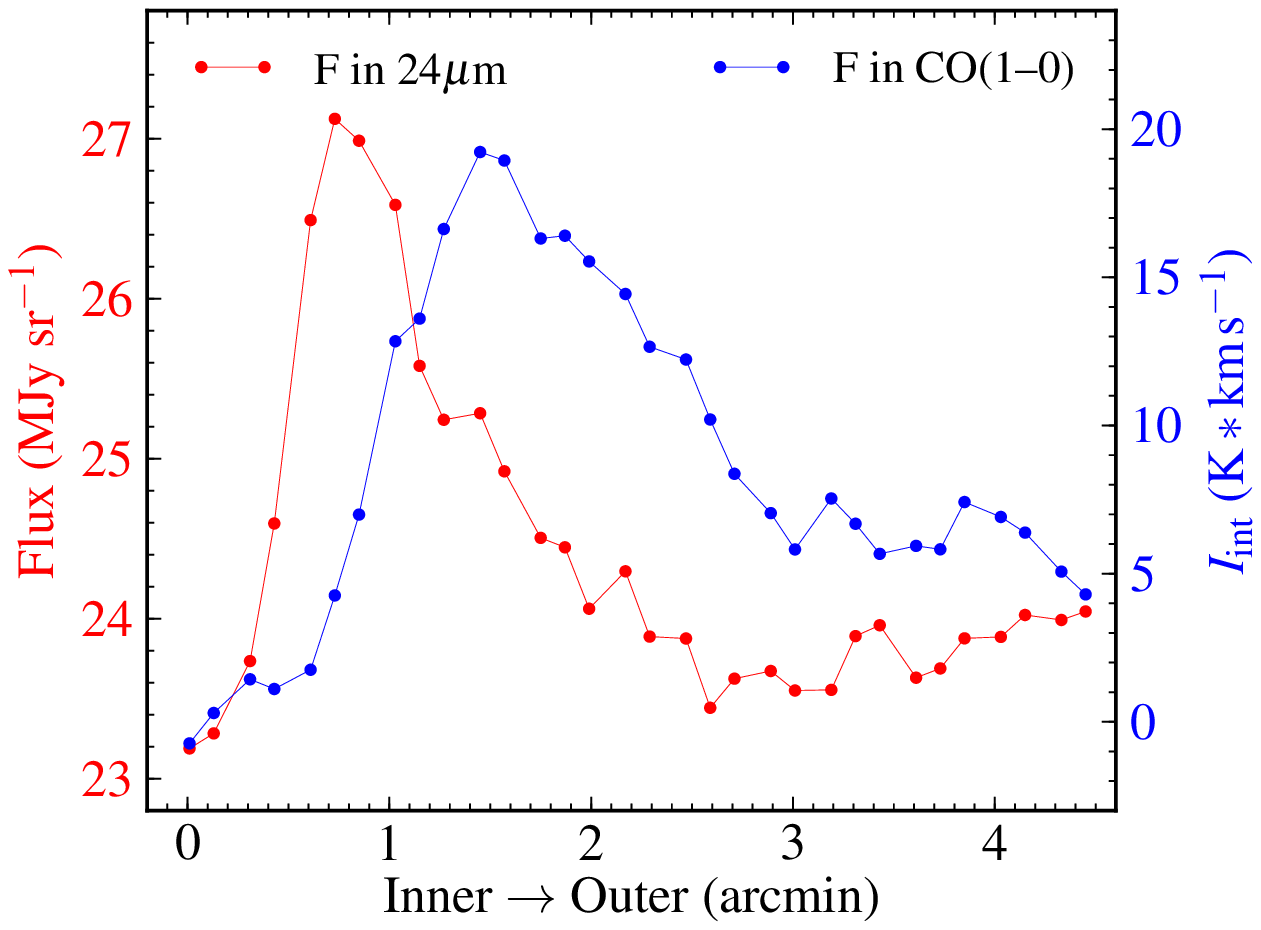}
\includegraphics[width=0.45\textwidth, angle=0]{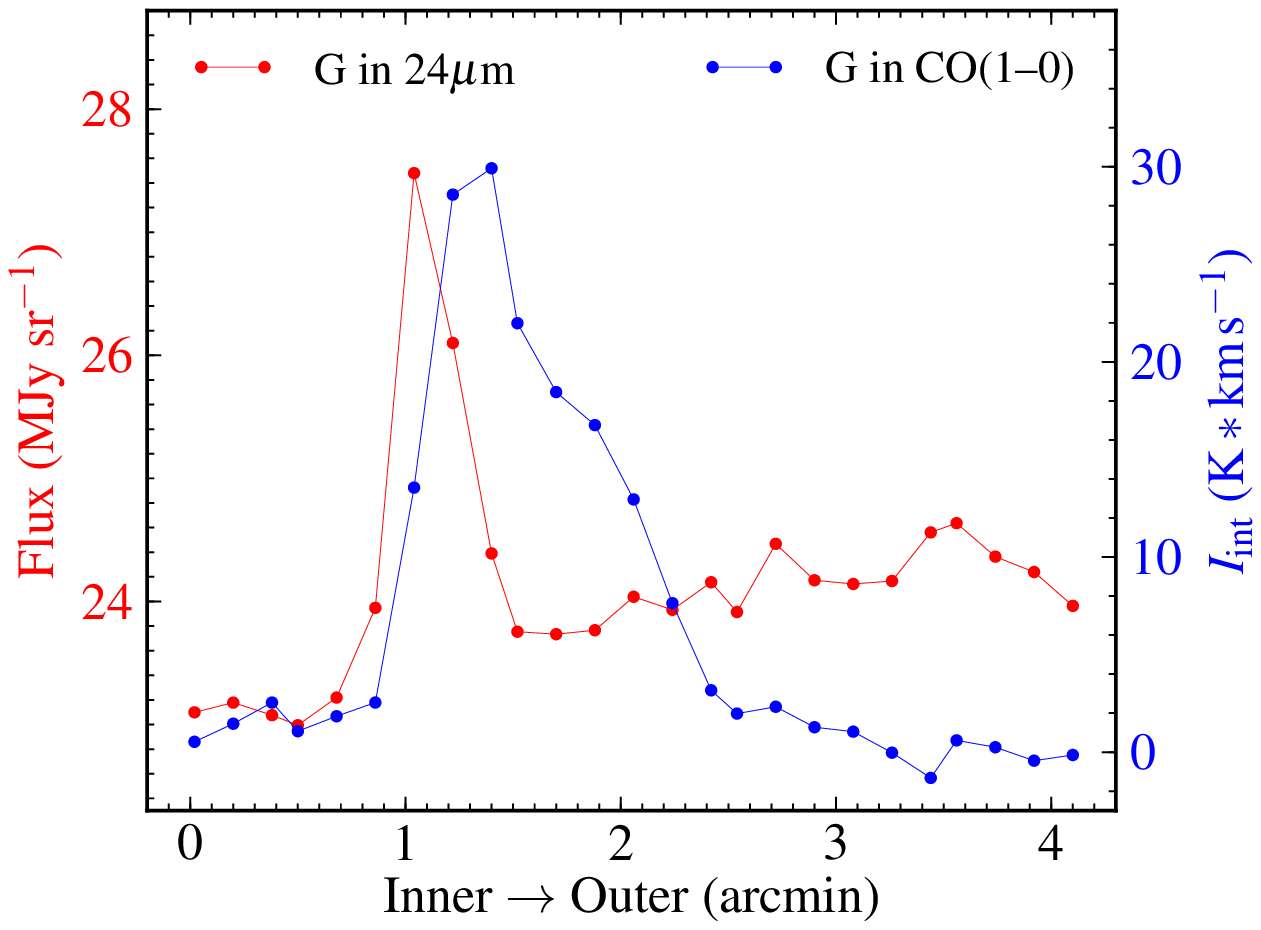}
\caption{Slice profiles of 24 $\mu$m (red) and CO (1-0) emission (blue) from the inner edge to the outer edge of the ring-like shell through the clumps AD, E, F, and G. The cutting paths are shown in Fig. \ref{Fig:int-map}. The profiles show a distribution of steep rise at the inner edge and a gradual fall at the outer edge. }
\label{Fig:slice}
\end{figure*}

\begin{figure*}
\centering
\includegraphics[width=0.45\textwidth, angle=0]{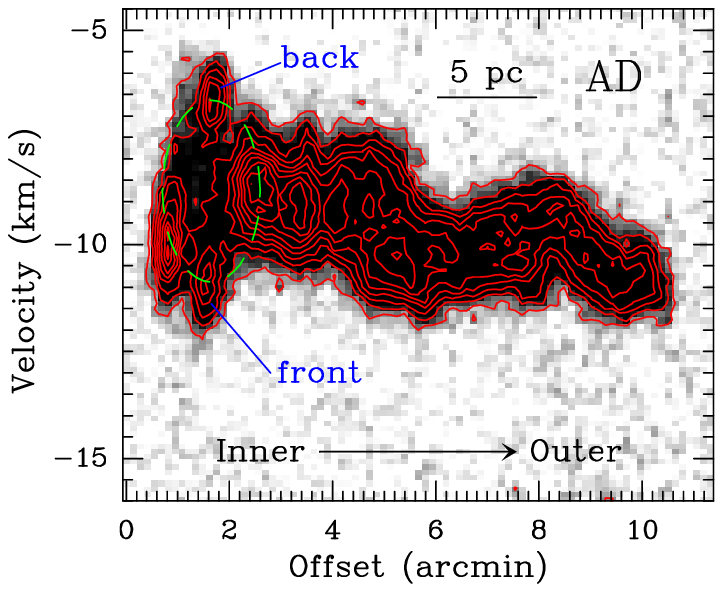}
\includegraphics[width=0.45\textwidth, angle=0]{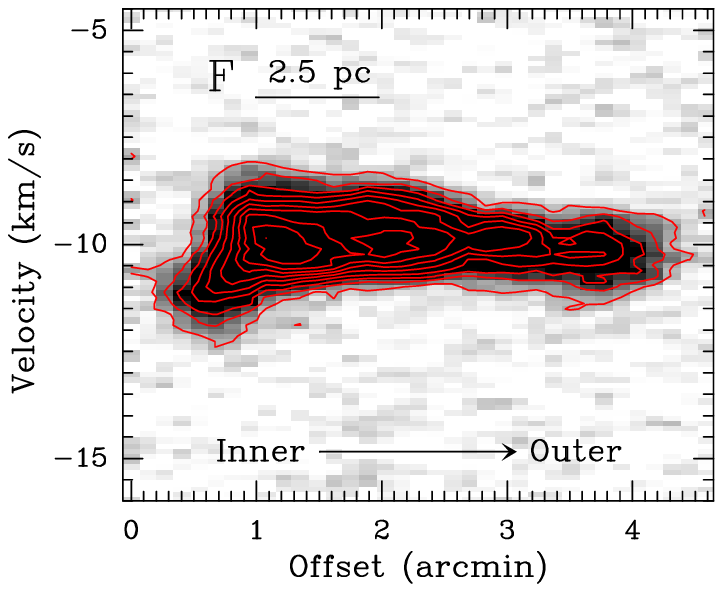}
\includegraphics[width=0.45\textwidth, angle=0]{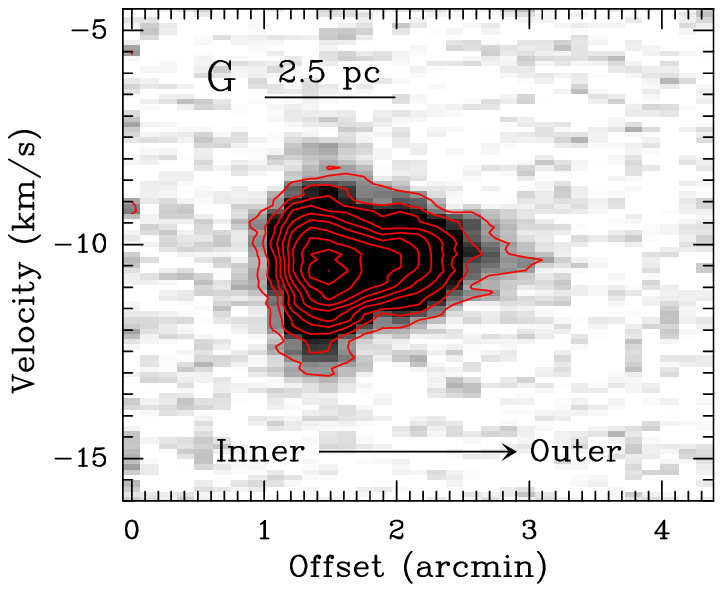}
\includegraphics[width=0.45\textwidth, angle=0]{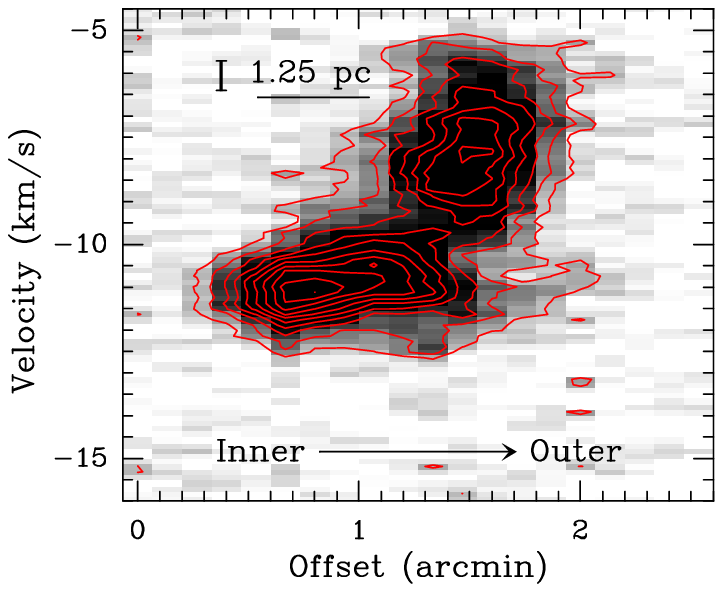}
\caption{Position-velocity diagram from the inner edge to the outer edge of the ring-like shell through the clumps AD, E, F, and I for the CO (1-0). The cutting paths are shown in Fig. \ref{Fig:int-map}. The green dashed ellipse indicates the 3D structure of the secondary bubble N131-A with front and back sides. Contour levels start at 3 times the rms level of 1.5 K and increase in steps of 1.5 K.} 
\label{Fig:pv}
\end{figure*}

In Fig. \ref{Fig:pv}, we show four CO (1-0) position-velocity (PV) diagrams for the cuts that we defined in Fig. \ref{Fig:slice}. The PV diagram can be used to identify a possible 3D structure of the bubble: an expanding shell should give rise to an arc-shaped structure in the PV space \citep{Peng2010}. The PV diagrams do not show obviously arc-shaped structures; however, the inner edge has larger velocity dispersion than the outer edge of the ring-like shell.

\begin{figure*}
\centering
\includegraphics[width=0.49\textwidth, angle=0]{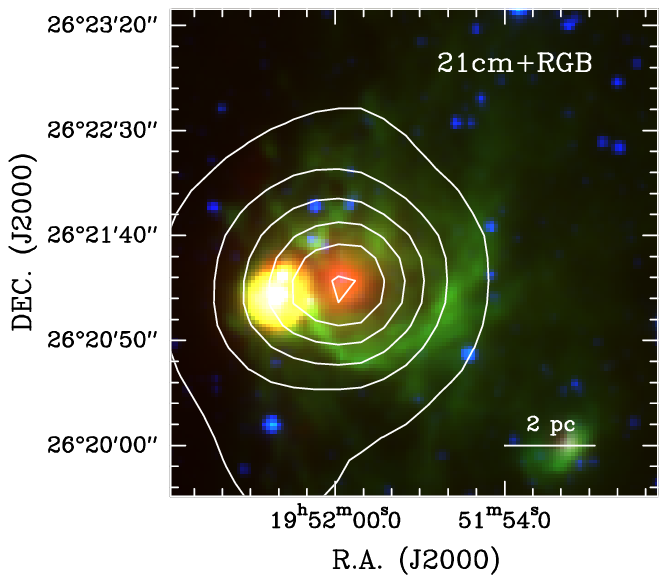}
\includegraphics[width=0.49\textwidth, angle=0]{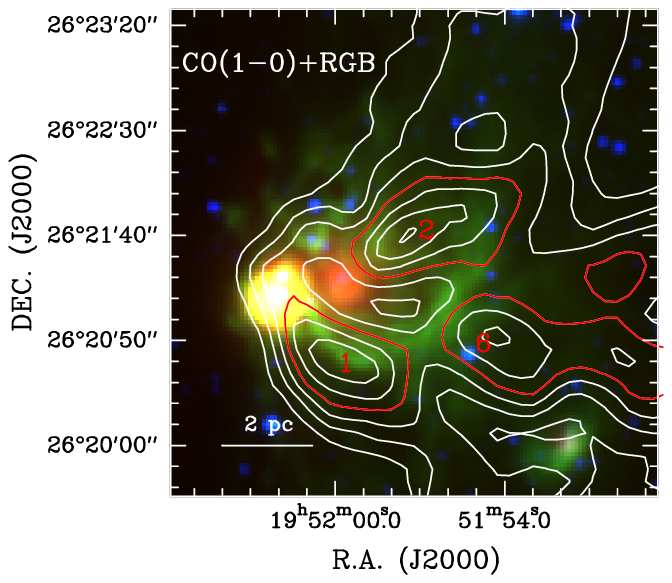}
\caption{ NVSS 21 cm continuum ($left$) and IRAM 30 m CO (1-0) ($right$) contours overlaid on a
Spitzer-IRAC three-colour image of the secondary bubble N131-A with 4.5 $\mu$m=blue, 8 $\mu$m=green, and 24 $\mu$m=red. The contour levels start at 5$\sigma$ in steps of 20$\sigma$ with $\sigma$ = 0.53 $\mjyb$ for the 21 cm, and at 5$\sigma$ in steps of 5$\sigma$ with $\sigma = 1.5\, {\rm K}(T^*_{\rm A})\, \kms$ for the CO (1-0). The numbers 1, 2, and 6 indicate fragment positions from Table \ref{tab_parameter}. The two maps above is from the zoom-in in the dashed square of Fig. \ref{Fig:irbubble}.}
\label{Fig:3color}
\end{figure*}

\begin{figure*}
\centering
\includegraphics[width=0.65\textwidth, angle=0]{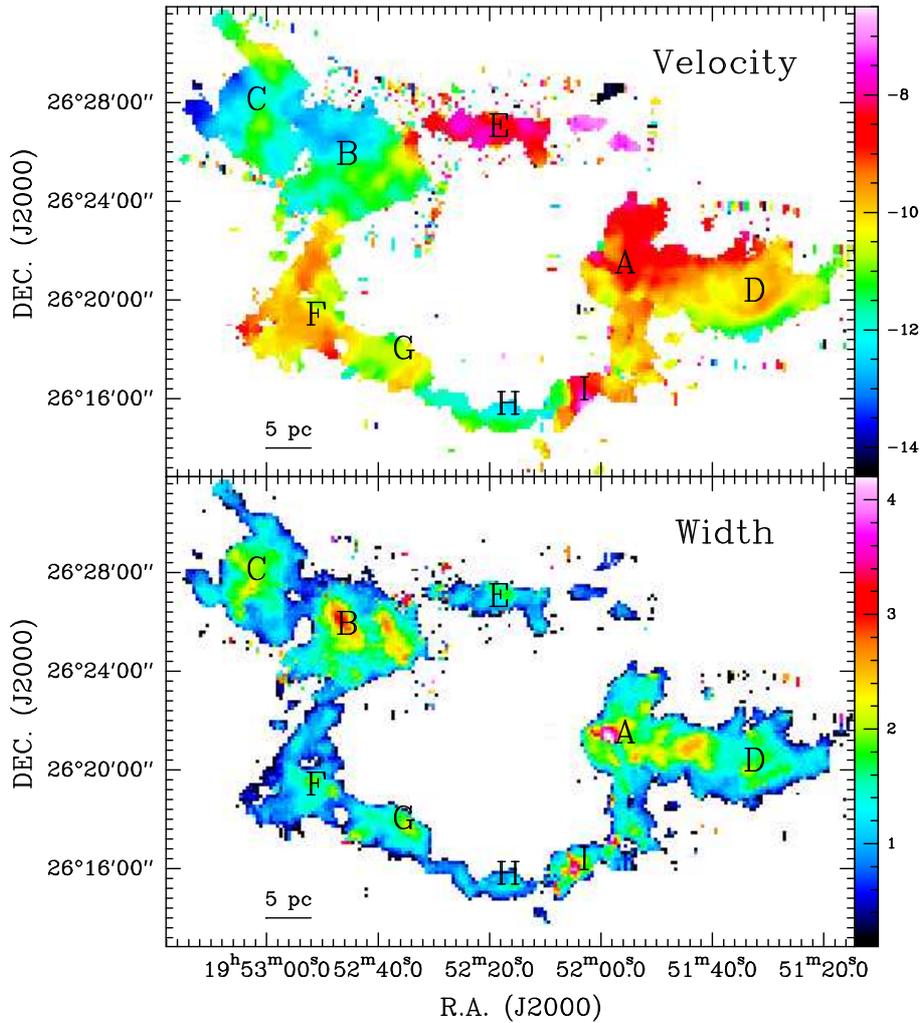}
\caption{The intensity-weighted mean velocity map (moment 1; $upper$) and velocity dispersion map (moment 2; $lower$). The unit of each colour bar is $\kms$. The letters from A to I indicate the positions of nine molecular clumps.}
\label{Fig:vel-width}
\end{figure*}

\begin{figure*}
\centering
\includegraphics[width=0.65\textwidth, angle=0]{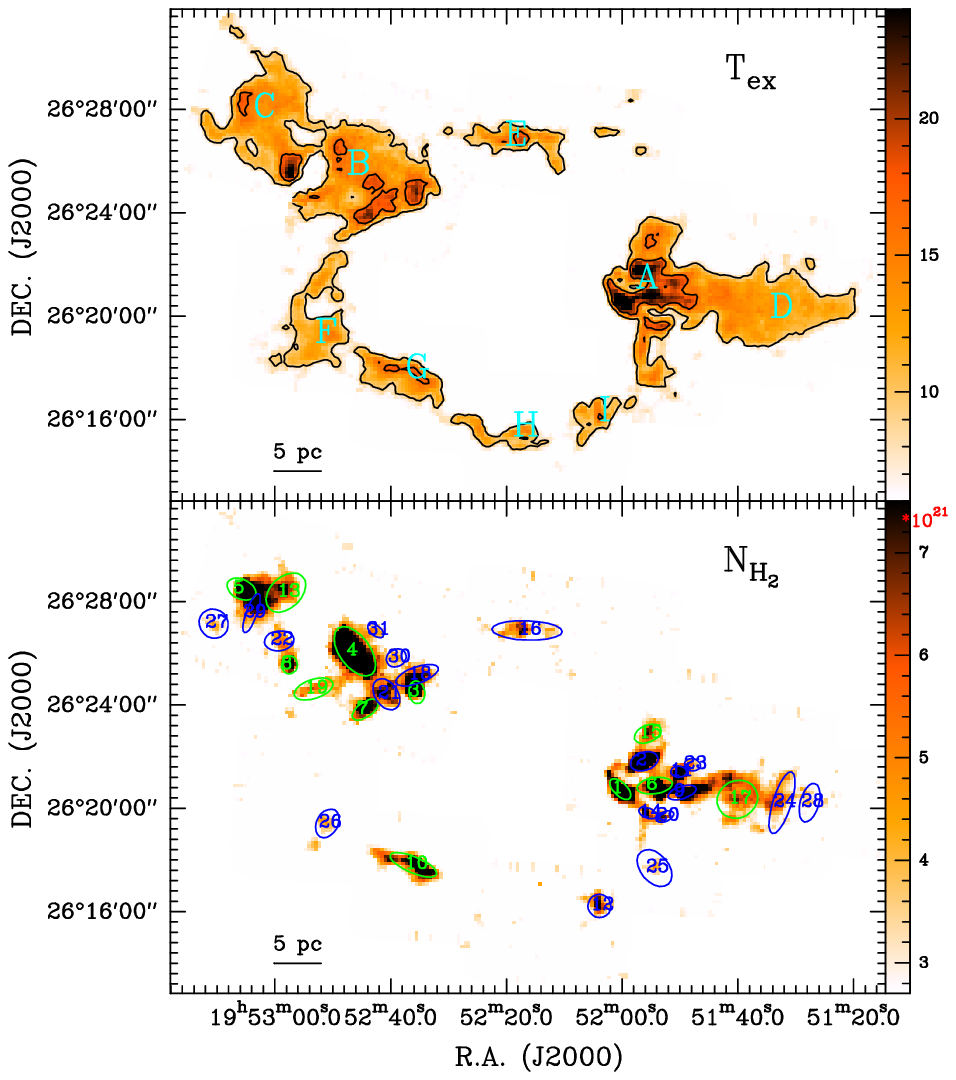}
\caption{N131: CO (1-0) excitation temperature map ($T_{\rm ex}$; $upper$) and H$_2$ column density distribution ($N_{\rm H_{2}}$; $lower$). For $T_{\rm ex}$, the contour levels are 9.1 and 16.5 K, and the maximum is 30.6 K. The ellipses with numbers indicate the position and size of each fragment  extracted by $Gaussclumps$. The letters from A to  I indicate the positions of nine molecular clumps. The green and blue ellipses correspond to to $\alpha_{\rm vir}<1$ and $\alpha_{\rm vir}>1$, respectively. The units for each colour bar are in Kelvin for $T_{\rm ex}$ and 10$^{21}$ cm$^{-2}$ for $N_{\rm H_{2}}$, respectively.}
\label{Fig:NH2-Tex}
\end{figure*}

In contrast, around clump A, the CO emission clearly exhibits a 3D structure with the front side and a back side of an expanding ring-like shell, which actually are the front and back sides of the secondary bubble N131-A (see details in Section \ref{sect:small_bubble}).  

\subsection{Secondary bubble N131-A}
\label{sect:small_bubble}
In the previous sections, we introduced a newly found secondary bubble N131-A centred at $\alpha$(J2000) = 19$\mathrm{^h}$52$\mathrm{^m}$00$\rlap.{^{\mathrm{s}}}$10 and $\delta$(J2000) = 26$^\circ$21$'$16$\dotsec$1, which is located at the position of  clump A (see the zoomed-in square in Fig. \ref{Fig:irbubble}). Figure \ref{Fig:3color} also shows the N131-A structure. The colour scale for each panel is a Spitzer-IRAC RGB image with 4.5 $\mu$m in blue, 8 $\mu$m in green, and 24 $\mu$m in red. The contours in the left and right panels are from NVSS 21 cm continuum and CO (1-0) emission, respectively. Seen from the left panel, the extended 8 $\mu$m is surrounding the compact 24 $\mu$m emission, and the peak of the 21 cm continuum is located at the peak of the compact 24 $\mu$m emission. Seen from the CO (1-0) emission three fragments, numbers 1, 2, and 6 (see Table \ref{tab_parameter} in Section \ref{sect:fragment_extraction}), shaped the shell of the secondary bubble N131-A. In Fig. \ref{Fig:pv}, the PV diagram through  clump A clearly shows a 3D structure with a front cloud and a back cloud. Its expanding speed is about 5.5 $\kms$. Therefore, the secondary bubble N131-A is a typical bubble with an expanding \HII region \citep{chur2006}.

Furthermore, we searched for the NVSS catalogue and obtained a total flux of $S_{\nu}$ = 152.3 mJy at $\nu$ = 1.4 GHz in the secondary bubble N131-A \citep{cond1998}. The flux of stellar Lyman photons $N_{\rm LyC}$ absorbed by the gas in the \HII region can be computed following the relation \citep{mezg1974}
\begin{equation}
    \label{eq:nlcy}
   \left(\frac{N_{\rm LyC}}{\rm s^{-1}}\right) = \frac{4.761 \times 10^{48}}{a(\nu, T_{\rm e})} \left(\frac{\nu}{\rm
GHz}\right)^{0.1} \left(\frac{T_{\rm e}}{\rm K}\right)^{-0.45} \left(\frac{S_{\nu}}{\rm Jy}\right) \left(\frac{D}{\rm
kpc}\right)^2,
\end{equation}
where $a(\nu, T_{\rm e}) \sim 1$ is a slowly varying function tabulated by \citet{mezg1967}, the effective temperature of the central star is assumed to be $T_{e} \sim$ 33000 K, and the distance is $D$ = 8.6 kpc. The power exponent of $T_{e}$ is small, so the result does not depend strongly on the chosen $T_{e}$. Based on the above, we derived a Lyman-continuum ionizing photon flux of log($N_{\rm LyC}$) = $\sim$ 47.7 from the \HII region, which is equivalent to an O9.5 star \citep{pana1973}.

\subsection{Moments 1 and 2}
The velocity structure (moment 1) of  bubble N131 has been analysed in Figs. 4, 5, and 6 of \citet{n131}, and in Figs. \ref{Fig:pv} and \ref{Fig:vel-width} of this work. Along the central axis of the clumps AD and BC, there is an apparent velocity gradient. At the systematic velocity of -10.5 $\kms$,  clump AD is redshifted with a velocity range of [-9.5 -6.5] $\kms$, while  clump BC is blueshifted with a velocity range of [-14.5 -11.5] $\kms$. This indicates that  clumps AD and BC are two filamentary flows transferring material outwardly, probably owing to the strong stellar wind. In addition, clumps G and H in the ring-like shell are blueshifted and  clumps E and I are redshifted, which indicates that the ring-like shell of the bubble is expanding outwardly.

For  clump D in Fig. \ref{Fig:vel-width}, in addition to  the velocity gradient in its east-west direction, there is another velocity gradient in its north-south direction. Clump F has a velocity component similar to the systematic velocity of -10.5 $\kms$. This can be explained if the expansion was blocked off by a pre-existing clump.

In the lower part of Fig. \ref{Fig:vel-width}, we present the velocity dispersion (moment 2) distribution of  bubble N131 determined from our CO (1-0) data. The maximum is 4.4 $\kms$ at the secondary bubble N131-A. The broadening is due to observing emission from both the front and the back sides along this sight line. At clumps B and I, the maximum is 3.4 and 4.0 $\kms$, respectively; at other clumps, the velocity dispersion ranges from 1.2 to 2.2 $\kms$. Overall, the velocity dispersion decreases from the inner edge to outer edge of the ring-like shell, which is consistent with the PV diagram in Fig. \ref{Fig:pv}.

\subsection{Excitation temperature and column density} 
\label{sect:column_desity}
Assuming that the lower transitions of the CO molecule are under local thermal equilibrium (LTE), their excitation temperature and column density can be computed \citep{winn1979,gard1991}. The CO (1-0) line is assumed to be optically thick, with a beam-filling factor of unity. Assuming that the excitation temperature and kinematic temperature for CO (1-0) are the same for $^{13}$CO (1-0), the excitation temperatures $T_{\rm ex}$ and then the column density $N_{\rm ^{13}CO}$ can be calculated directly using conversion factors of $\rm [H_2]/[^{12}CO] = 10^4$ and $\rm [CO]/[^{13}CO]=60$. The obtained $T_{\rm ex}$ and $N_{\rm H_2}$ maps are shown in Fig. \ref{Fig:NH2-Tex}.

In the upper part of Fig. \ref{Fig:NH2-Tex}, $T_{\rm ex}$ mainly ranges from 9.1 to 16.5 K in the diffuse structure, while excitation temperatures between 16.5 and 30.6 K are found towards several compact fragments. The higher $T_{\rm ex}$ values are generally found at the inner edge of the ring-like shell whose temperature decreases from the inside to the outside. We suggest that the heating is caused by warm stellar winds within the bubble.

In the lower part of Fig. \ref{Fig:NH2-Tex}, $N_{\rm H_2}$ mainly ranges from $4.0\times 10^{21}$ to $14.0\times 10^{21}$ cm$^{-2}$, and the maximum is $21.4\times 10^{21}$ cm$^{-2}$ located at fragment No.1 (see Table \ref{tab_parameter}) in clump A. The denser regions are distributed at  clumps AD and BC, while the ring-like shell of the bubble N131 is very diffuse and just has few dense fragments. Furthermore,  clump A is denser than clump D, and  clump B is as dense as clump C. These clumps have fragmented into several dense fragments, which is  discussed in Section \ref{sect:fragmentation}.

The $^{13}$CO isotope probes only relatively low volume densities, and may be depleted in fragments with high densities. It will lead to underestimating the densities and masses of the fragments. Alternatively, the dust continuum emission can be used to derive column densities, but based on the CO observations, there are several velocity components found on the  line of sight towards the N131 which continuum data cannot distinguish. In Fig. \ref{Fig:irbubble} it is obvious that the Herschel data is  in general not well correlated with the $^{13}$CO emission. However, several of the fragments show clear counterparts in the Herschel images, and so we combined 70, 160, and 250 $\mu$m data to estimate the column densities \citep{Preibisch2012}, which are found to be comparable to the column densities from $^{13}$CO emission. In the following, we therefore use the column densities and masses derived from $^{13}$CO.

\section{Discussion} 
\label{sec:discussion}

 \begin{figure*}
\centering
\includegraphics[width=0.45\textwidth, angle=0]{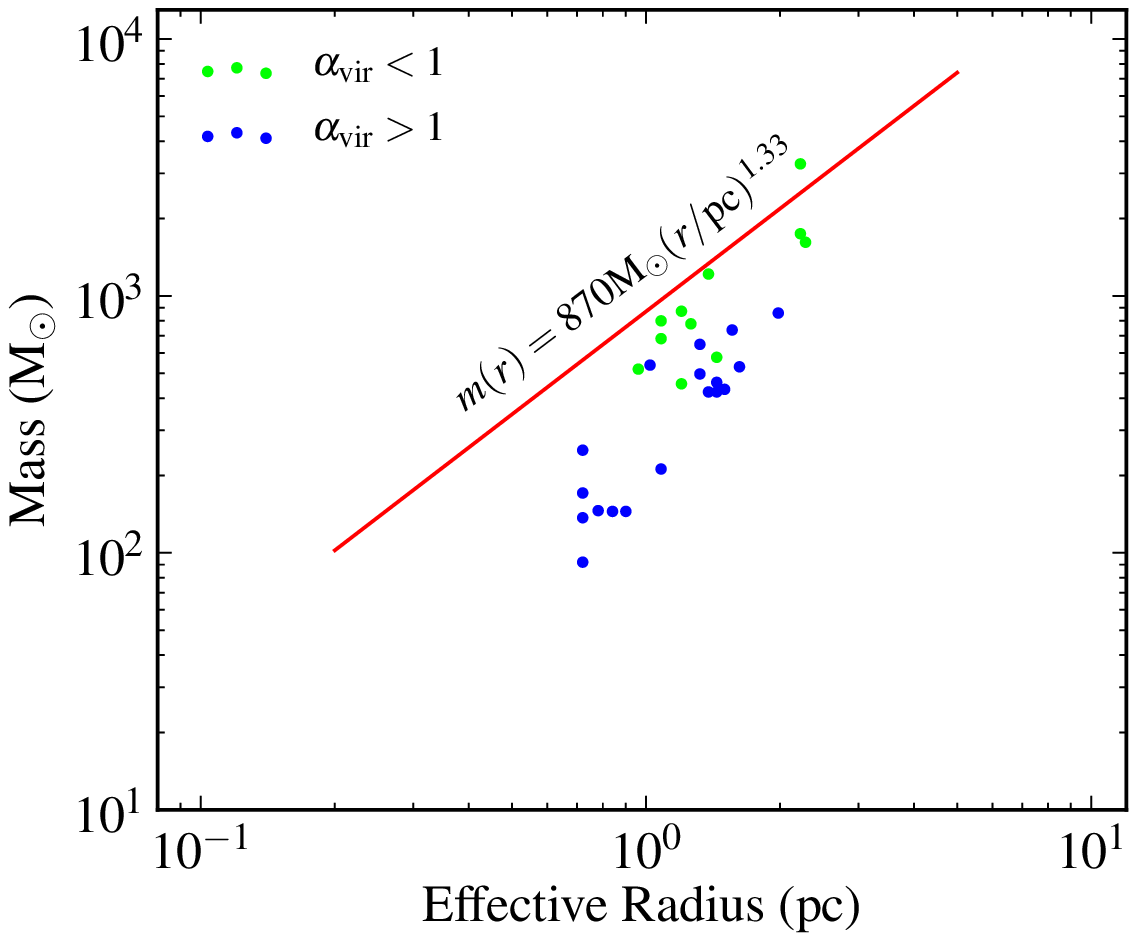}
\caption{$Mass$-$radius$ distributions of Gaussian fragments extracted from $Gaussclumps$. The masses are listed in Table \ref{tab_parameter} and the effective radius is derived with $r = FWHM / (2\sqrt{\rm ln2})$. The red line delineates the threshold introduced by \citet{Kauffmann2010} separating the regimes under which high-mass stars can form (above the line) or not (below the line). The green and blue points  belong to $\alpha_{\rm vir}<1$ and $\alpha_{\rm vir}>1$, respectively. }
\label{Fig:mass_size}
\end{figure*}

\subsection{Stellar wind}
The strong stellar winds from the central OB-type stars, along with overpressure from ionization and high temperature, may cause the expansion of gas/dust shells in their vicinity, of which N131 is an example \citep{Castor1975,Weaver1977}. In particular, dust emission at 24 $\mu$m can be used as an indicator of the effects of winds. Very small grains, heated by the stellar radiation and out of thermal equilibrium, radiate at 24 $\mu$m \citep{Jones1999}. \citet{deha2010} found that 86\% of the bubbles contain ionized gas detected in 20 cm radio continuum radiation and 98\% of the bubbles exhibit extended and dominated 24 $\mu$m emission in their central regions. A central hole in 24 $\mu$m emission  maps of  some \HII regions might  be  an  indication  that  winds  are  at  work \citep{wats2008}.  Such a hole could  also  be  attributed  to  the  radiation pressure of the central ionizing stars or to dust destruction by the stars’ ionizing radiation \citep{krum2009,Martins2010} and strong stellar winds \citep{Castor1975,Weaver1977}.

\begin{figure*}
\centering
\includegraphics[width=0.55\textwidth, angle=0]{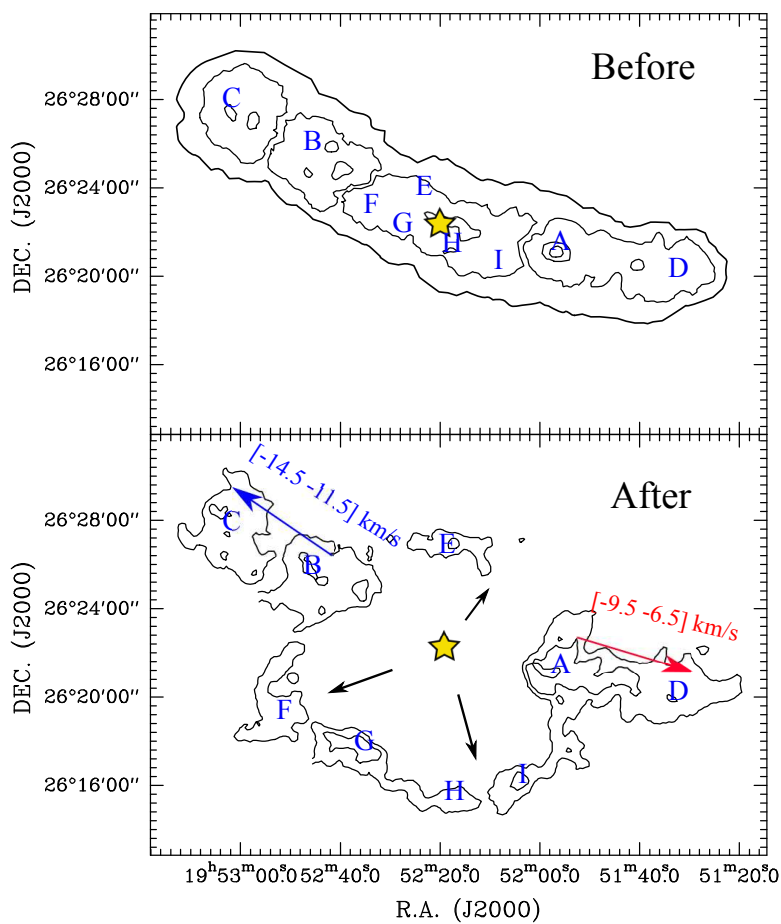}
\caption{A schematic illustration depicting our target region before and after the formation of  bubble N131. The yellow star indicates a group of ionizing stars that caused the expansion of the shell. In the upper panel, we indicate the proposed pre-existing clumps A to I. In the lower panel, we indicate the observed clumps A to I and the velocity structure.}
\label{Fig:before-after}
\end{figure*}

In the direction of the centre of bubble N131, we detect very weak 21 cm continuum and 24 $\mu$m emission; it has been suggested that the emission is an unrelated background component, so bubble N131 is a rare sample in a different evolutionary stage \citep{deha2010}. However, strong emission at 24 $\mu$m was observed in the direction of the dense PDR filaments delineating the ring-like shell of the bubble. The 24 $\mu$m emission mostly originates in the PDR filament, and may have been compressed by the strong stellar winds emitted by the ionizing stars. This indicates that the material giving rise to the 24 $\mu$m emission inside N131 cannot survive the intense radiation field produced by the central cluster, and the bubble is likely evacuated of ionized gas and hot dust by the strong stellar winds of early OB-type stars. Therefore,  bubble N131 is at an advanced stage of evolution.

\subsection{Fragmentation}
\label{sect:fragmentation}

In the lower panel of Fig. \ref{Fig:NH2-Tex}, we show the Gaussian fragments extracted using \textit{RUN~GAUSSCLUMPS} in the GILDAS software package. The position and size of each fragment is indicated with an ellipse in the column density map of Fig. \ref{Fig:NH2-Tex}. These fragments are mostly distributed at  clumps AD and BC, and there are only five fragments located at the other clumps. This suggests that the ring-like shell is more diffuse than  clumps AD and BC, and fragmentation is more pronounced towards clumps AD and BC than in the ring-like shell.

In Fig. \ref{Fig:mass_size}, we present the mass-size relation diagram for the extracted fragments. Comparison with the high-mass star formation threshold of $m(r) > 870 {\rm \Msun} (r/{\rm pc})^{1.33}$ empirically proposed by \citet{Kauffmann2010} allows us to determine whether these fragments are capable of giving birth to massive stars. The data points are distributed below the threshold (given by the red line in Fig. \ref{Fig:mass_size}) that discriminates between high- and low-mass star formation whose entries fall above and below the line, respectively, indicative of low-mass star-forming candidates. \citet{Lee2007} suggested that radiation from massive stars can only trigger the formation of low- and intermediate-mass objects. It appears that the stellar winds in the N131 bubble only triggered  the formation of low-mass stars. It is also likely that some of these fragments in the clumps AD and BC were pre-existing and were not triggered by the stellar winds.

\subsection{Virial state}

\begin{figure*}
\centering
\includegraphics[width=0.45\textwidth, angle=0]{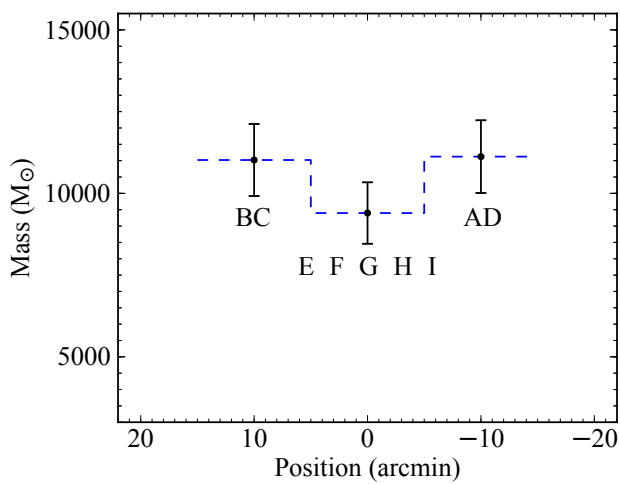}
\caption{$Mass$-$position$ distributions of the clumps A to I in  bubble N131. The position axis  of the labelled clumps is offset from the bubble centre. Filaments AD, BC, and the pre-existing clump AB have a similar projected length of about 10 arcmin. Filament BC; clumps E, F, G, H, I (i.e. the ring-like shell); and filament AD have a mass of 11000, 9000, and 11000 $\Msun$, respectively.  The error bar is 10\% for each mass value.}
\label{Fig:mass_length}
\end{figure*}

The virial theorem provides a method to test whether a molecular fragment is in a stable state. If external pressure and magnetic fields are ignored, $M_{\rm vir} \simeq 210\, r\, \Delta V^{2}\, (\Msun)$ \citep{Evans1999}, where $r$ is the fragment radius in pc and $\Delta V^{2}$ is the full width at half-maximum linewidth in $\kms$. Between the virial mass $M_{\rm vir}$ and the fragment mass $M$, if the virial parameter $\alpha_{\rm vir} = M_{\rm vir}/M < 1$, the molecular fragment is gravitationally bound, potentially unstable, and collapsing; if $\alpha_{\rm vir} > 1$, the fragment is not gravitationally bound in a stable or expanding state \citep{Hindson2013}.

\citet{Kauffmann2010} found empirically a massive star formation threshold $m(r) = 870 {\Msun} (r/{\rm pc})^{1.33}$ for solar neighbourhood clouds. With our velocity-resolved observations, we address its physical origin studying the virial states of the fragments.

The derived virial masses $M_{\rm vir}$ and virial parameters $\alpha_{\rm vir}$ are listed in Table \ref{tab_parameter}. We also show entries for the corresponding fragments with $\alpha_{\rm vir} < 1$ (in green) and $\alpha_{\rm vir} > 1$ (in blue) in Figs. \ref{Fig:NH2-Tex} and \ref{Fig:mass_size}. Of the extracted 31 fragments, 12 fragments are gravitationally bound, and 19 fragments are not gravitationally bound. In Fig. \ref{Fig:mass_size}, the gravitationally bound and unbound fragments are distributed differently. The gravitationally bound fragments ($\alpha_{\rm vir} < 1$) tend to follow the mass-size relation ($m(r) = 870 {\Msun} (r/{\rm pc})^{1.33}$) better, and this implies a gravitational origin of the relation. This favours a scenario where only the fragments that are gravitationally bound can further contract to the limit $m(r) = 870 {\Msun} (r/{\rm pc})^{1.33}$.

\subsection{How to form  bubble N131?}

Based on the intriguing morphological structure and velocity distribution of  N131 presented above, we propose that the progenitor of  bubble N131 was a filamentary nebula elongated along the direction of the clumps AD and BC (see the sketch in Fig. \ref{Fig:before-after}). It is likely that within the middle of the filamentary nebula, a cluster of massive stars ionized, heated, and blew away the surrounding interstellar medium and compressed the pre-existing clumps. With the expansion of \HII region at relatively high temperature, pressure, and  strong stellar winds lasting more than 10$^6$ yr, the feedback from the massive stars broke up the pre-existing filamentary nebula and separated it into the two unconnected clumps AD and BC. With the expansion of evolving \HII region, a ring-like structure is formed. \citet{Bodenheimer1979} have simulated a similar scenario. We provide further evidence to back up this argument.

Clumps AD and BC are two unconnected filamentary nebula on the opposite sides of the bubble. Along the direction of  clumps C, B, A, and D in Fig. \ref{Fig:int-map}, there is a velocity gradient from -14.5 to -6.5 $\kms$ \citep{n131}. This velocity coherence indicates that in the past  clumps AD and BC were likely connected together. Therefore, the velocity gradient may be partly produced by the bubble expansion.

These broken filamentary morphologies, caused by bubble expansion, are rather common in large-scale surveys, i.e. the 500 pc filamentary gas wisp from \citet{Li2013} and the bubble N39 from \citet{deha2010}. In addition, the secondary bubble N131-A (see Section \ref{sect:small_bubble}) close to  N131 is also embedded in the filamentary nebula (or clumps AD); N131-A is expanding outwardly, and compressing the surrounded material. Bubble N131 in its early stage might have had a similar evolutionary process as  N131-A. With the expansion of the bubble,  the strong stellar winds finally help to form bubble N131.

To test our scenario, we also compared the relative masses of these clumps surrounding bubble N131 with the proposed pre-existing clumps in Fig. \ref{Fig:mass_length}. These clumps are divided into three parts: clump AD, clump BC, and the ring-like shell (mainly clumps E, F, G, H, and I). The masses of  clumps E, F, G, H, and I in the ring-like shell are mainly from the proposed pre-existing clump AB.  Considering that clumps AD, BC, and AB have a similar projection area, we assumed that  clumps AD, BC, and AB had the same mass when it was a connected filamentary nebula in the past.  The mass distributions of the three parts are shown in Fig. \ref{Fig:mass_length}. Clump AD has a very similar mass to  clump BC, while the mass of the ring-like shell is a little lower (about 16\%) than clumps AD and BC. The 16\% reduction  can be explained as some masses from the assumed filamentary clump AB being merged into  clumps A and B. Above all, the similarity of the  masses of  clump AD, clump BC, and the ring-like shell suggests that the bubble might originate from a filamentary nebula if mass conserves.

\section{Summary}
\label{sect:summary}
Bubble N131 shows an intriguing morphology and hierarchical velocity distribution, which have been reported in the context of previous CO observations. However, the angular resolution was low, and the bubble formation is still unclear.

In this work, we analysed archival multiwavelength observations including 3.6, 4.5, 5.8, 8.0, 24, 70, 160, 250, 350, 500 $\mu$m, 1.1 mm, and 21 cm towards  bubble N131. These observations revealed the distributions of the warm and cold dust emission. Bubble N131 is a special object owing to its large hole at 24 $\mu$m and 21 cm in the direction of the bubble centre. Comparing the 4.5 and 8.0 $\mu$m with the 24 $\mu$m and 21 cm, we found a secondary bubble N131-A located at the filamentary clump AD. The derived Lyman-continuum ionizing photon flux within the N131-A is equivalent to an O9.5 star.

We performed new observations of CO (2-1), CO (1-0), and $^{13}$CO (1-0) using the IRAM 30 m telescope. The distribution of molecular gas is largely consistent with  the dust emission from  5.8 $\mu$m to 500 $\mu$m. The dust and gas clumps are compact and might have been compressed at the inner edge of the ring-like shell, while they are extended and might be pre-existing at the outer edge. The column density, excitation temperature, and velocity also show a potentially hierarchical distribution from the inner to outer edges of the ring-like shell. We detected the front and back sides of the secondary bubble N131-A  in the direction of its centre. Based on the above, we propose a scenario in which  bubble N131 was formed by the expansion of \HII region and strong stellar winds from a group of massive stars inside the bubble.

Comparing the mass-size relation with high-mass formation threshold $m(r) = 870 {\Msun} (r/{\rm pc})^{1.33}$, the bubble is fragmenting into dense fragments which can potentially form low-mass stars. These molecular clumps fragmented into 31 fragments, most of which are located at the two giant filamentary nebula (or the clumps AD and BC). Twelve of the fragments are gravitationally bound and 19 are gravitationally unbound. From the morphology, infrared dust emission, velocity dispersion, and mass distribution, bubble N131 likely originated in a filamentary nebula, within which the strong stellar winds from a group of massive stars broke up a pre-existing filamentary nebula into  clumps AD and BC, and swept up the surrounded material into a ring-like shell of  bubble N131.

\begin{acknowledgements}
We wish to thank the anonymous referee for comments and suggestions that improved the clarity of the paper. C.-P. Zhang acknowledges support by the MPG-CAS Joint Doctoral Promotion Program (DPP) and observations by the IRAM staff. This work is partly supported by the National Key Basic Research Program of China (973 Program) 2015CB857100 and National Natural Science Foundation of China 11363004 and 11403042. 

\end{acknowledgements}

\bibliographystyle{aa}
\bibliography{references}

\begin{thebibliography}{60}
\expandafter\ifx\csname natexlab\endcsname\relax\def\natexlab#1{#1}\fi

\bibitem[{{Aguirre} {et~al.}(2011){Aguirre}, {Ginsburg}, {Dunham}, {Drosback},
  {Bally}, {Battersby}, {Bradley}, {Cyganowski}, {Dowell}, {Evans}, {Glenn},
  {Harvey}, {Rosolowsky}, {Stringfellow}, {Walawender}, \&
  {Williams}}]{Aguirre2011}
{Aguirre}, J.~E., {Ginsburg}, A.~G., {Dunham}, M.~K., {et~al.} 2011, \apjs,
  192, 4

\bibitem[{{Anderson} {et~al.}(2012){Anderson}, {Zavagno}, {Deharveng},
  {Abergel}, {Motte}, {Andr{\'e}}, {Bernard}, {Bontemps}, {Hennemann}, {Hill},
  {Rod{\'o}n}, {Roussel}, \& {Russeil}}]{Anderson2012}
{Anderson}, L.~D., {Zavagno}, A., {Deharveng}, L., {et~al.} 2012, \aap, 542,
  A10

\bibitem[{{Beaumont} \& {Williams}(2010)}]{beau2010}
{Beaumont}, C.~N. \& {Williams}, J.~P. 2010, \apj, 709, 791

\bibitem[{{Benjamin} {et~al.}(2003){Benjamin}, {Churchwell}, {Babler}, {Bania},
  {Clemens}, {Cohen}, {Dickey}, {Indebetouw}, {Jackson}, {Kobulnicky},
  {Lazarian}, {Marston}, {Mathis}, {Meade}, {Seager}, {Stolovy}, {Watson},
  {Whitney}, {Wolff}, \& {Wolfire}}]{benj2003}
{Benjamin}, R.~A., {Churchwell}, E., {Babler}, B.~L., {et~al.} 2003, \pasp,
  115, 953

\bibitem[{{Bodenheimer} {et~al.}(1979){Bodenheimer}, {Tenorio-Tagle}, \&
  {Yorke}}]{Bodenheimer1979}
{Bodenheimer}, P., {Tenorio-Tagle}, G., \& {Yorke}, H.~W. 1979, \apj, 233, 85

\bibitem[{{Cappa} {et~al.}(2014){Cappa}, {Rubio}, {Romero}, {Duronea}, \&
  {Firpo}}]{Cappa2014}
{Cappa}, C.~E., {Rubio}, M., {Romero}, G.~A., {Duronea}, N.~U., \& {Firpo}, V.
  2014, \aap, 562, A6

\bibitem[{{Carey} {et~al.}(2009){Carey}, {Noriega-Crespo}, {Mizuno}, {Shenoy},
  {Paladini}, {Kraemer}, {Price}, {Flagey}, {Ryan}, {Ingalls}, {Kuchar},
  {Pinheiro Gon{\c c}alves}, {Indebetouw}, {Billot}, {Marleau}, {Padgett},
  {Rebull}, {Bressert}, {Ali}, {Molinari}, {Martin}, {Berriman}, {Boulanger},
  {Latter}, {Miville-Deschenes}, {Shipman}, \& {Testi}}]{care2009}
{Carey}, S.~J., {Noriega-Crespo}, A., {Mizuno}, D.~R., {et~al.} 2009, \pasp,
  121, 76

\bibitem[{{Castor} {et~al.}(1975){Castor}, {McCray}, \& {Weaver}}]{Castor1975}
{Castor}, J., {McCray}, R., \& {Weaver}, R. 1975, \apjl, 200, L107

\bibitem[{{Churchwell} {et~al.}(2009){Churchwell}, {Babler}, {Meade},
  {Whitney}, {Benjamin}, {Indebetouw}, {Cyganowski}, {Robitaille}, {Povich},
  {Watson}, \& {Bracker}}]{chur2009}
{Churchwell}, E., {Babler}, B.~L., {Meade}, M.~R., {et~al.} 2009, \pasp, 121,
  213

\bibitem[{{Churchwell} {et~al.}(2006){Churchwell}, {Povich}, {Allen}, {Taylor},
  {Meade}, {Babler}, {Indebetouw}, {Watson}, {Whitney}, {Wolfire}, {Bania},
  {Benjamin}, {Clemens}, {Cohen}, {Cyganowski}, {Jackson}, {Kobulnicky},
  {Mathis}, {Mercer}, {Stolovy}, {Uzpen}, {Watson}, \& {Wolff}}]{chur2006}
{Churchwell}, E., {Povich}, M.~S., {Allen}, D., {et~al.} 2006, \apj, 649, 759

\bibitem[{{Churchwell} {et~al.}(2007){Churchwell}, {Watson}, {Povich},
  {Taylor}, {Babler}, {Meade}, {Benjamin}, {Indebetouw}, \&
  {Whitney}}]{chur2007}
{Churchwell}, E., {Watson}, D.~F., {Povich}, M.~S., {et~al.} 2007, \apj, 670,
  428

\bibitem[{{Condon} {et~al.}(1998){Condon}, {Cotton}, {Greisen}, {Yin},
  {Perley}, {Taylor}, \& {Broderick}}]{cond1998}
{Condon}, J.~J., {Cotton}, W.~D., {Greisen}, E.~W., {et~al.} 1998, \aj, 115,
  1693

\bibitem[{{Csengeri} {et~al.}(2011){Csengeri}, {Bontemps}, {Schneider},
  {Motte}, \& {Dib}}]{Csengeri2011}
{Csengeri}, T., {Bontemps}, S., {Schneider}, N., {Motte}, F., \& {Dib}, S.
  2011, \aap, 527, A135

\bibitem[{{Deharveng} {et~al.}(2010){Deharveng}, {Schuller}, {Anderson},
  {Zavagno}, {Wyrowski}, {Menten}, {Bronfman}, {Testi}, {Walmsley}, \&
  {Wienen}}]{deha2010}
{Deharveng}, L., {Schuller}, F., {Anderson}, L.~D., {et~al.} 2010, \aap, 523,
  A6

\bibitem[{{Dyson} \& {Williams}(1980)}]{Dyson1980}
{Dyson}, J.~E. \& {Williams}, D.~A. 1980, {Physics of the interstellar medium}

\bibitem[{{Evans}(1999)}]{Evans1999}
{Evans}, II, N.~J. 1999, \araa, 37, 311

\bibitem[{{Faimali} {et~al.}(2012){Faimali}, {Thompson}, {Hindson}, {Urquhart},
  {Pestalozzi}, {Carey}, {Shenoy}, {Veneziani}, {Molinari}, \&
  {Clark}}]{Faimali2012}
{Faimali}, A., {Thompson}, M.~A., {Hindson}, L., {et~al.} 2012, \mnras, 426,
  402

\bibitem[{{Garden} {et~al.}(1991){Garden}, {Hayashi}, {Hasegawa}, {Gatley}, \&
  {Kaifu}}]{gard1991}
{Garden}, R.~P., {Hayashi}, M., {Hasegawa}, T., {Gatley}, I., \& {Kaifu}, N.
  1991, \apj, 374, 540

\bibitem[{{Ginsburg} {et~al.}(2013){Ginsburg}, {Glenn}, {Rosolowsky},
  {Ellsworth-Bowers}, {Battersby}, {Dunham}, {Merello}, {Shirley}, {Bally},
  {Evans}, {Stringfellow}, \& {Aguirre}}]{Ginsburg2013}
{Ginsburg}, A., {Glenn}, J., {Rosolowsky}, E., {et~al.} 2013, \apjs, 208, 14

\bibitem[{{G{\'o}mez} {et~al.}(2011){G{\'o}mez}, {Wyrowski}, {Pillai},
  {Leurini}, \& {Menten}}]{Gomez2011}
{G{\'o}mez}, L., {Wyrowski}, F., {Pillai}, T., {Leurini}, S., \& {Menten},
  K.~M. 2011, \aap, 529, A161

\bibitem[{{Goodman} {et~al.}(2014){Goodman}, {Alves}, {Beaumont}, {Benjamin},
  {Borkin}, {Burkert}, {Dame}, {Jackson}, {Kauffmann}, {Robitaille}, \&
  {Smith}}]{Goodman2014}
{Goodman}, A.~A., {Alves}, J., {Beaumont}, C.~N., {et~al.} 2014, \apj, 797, 53

\bibitem[{{Griffin} {et~al.}(2010){Griffin}, {Abergel}, {Abreu}, {Ade},
  {Andr{\'e}}, {Augueres}, {Babbedge}, {Bae}, {Baillie}, {Baluteau}, {Barlow},
  {Bendo}, {Benielli}, {Bock}, {Bonhomme}, {Brisbin}, {Brockley-Blatt},
  {Caldwell}, {Cara}, {Castro-Rodriguez}, {Cerulli}, {Chanial}, {Chen},
  {Clark}, {Clements}, {Clerc}, {Coker}, {Communal}, {Conversi}, {Cox},
  {Crumb}, {Cunningham}, {Daly}, {Davis}, {de Antoni}, {Delderfield}, {Devin},
  {di Giorgio}, {Didschuns}, {Dohlen}, {Donati}, {Dowell}, {Dowell}, {Duband},
  {Dumaye}, {Emery}, {Ferlet}, {Ferrand}, {Fontignie}, {Fox}, {Franceschini},
  {Frerking}, {Fulton}, {Garcia}, {Gastaud}, {Gear}, {Glenn}, {Goizel},
  {Griffin}, {Grundy}, {Guest}, {Guillemet}, {Hargrave}, {Harwit}, {Hastings},
  {Hatziminaoglou}, {Herman}, {Hinde}, {Hristov}, {Huang}, {Imhof}, {Isaak},
  {Israelsson}, {Ivison}, {Jennings}, {Kiernan}, {King}, {Lange}, {Latter},
  {Laurent}, {Laurent}, {Leeks}, {Lellouch}, {Levenson}, {Li}, {Li},
  {Lilienthal}, {Lim}, {Liu}, {Lu}, {Madden}, {Mainetti}, {Marliani}, {McKay},
  {Mercier}, {Molinari}, {Morris}, {Moseley}, {Mulder}, {Mur}, {Naylor},
  {Nguyen}, {O'Halloran}, {Oliver}, {Olofsson}, {Olofsson}, {Orfei}, {Page},
  {Pain}, {Panuzzo}, {Papageorgiou}, {Parks}, {Parr-Burman}, {Pearce},
  {Pearson}, {P{\'e}rez-Fournon}, {Pinsard}, {Pisano}, {Podosek}, {Pohlen},
  {Polehampton}, {Pouliquen}, {Rigopoulou}, {Rizzo}, {Roseboom}, {Roussel},
  {Rowan-Robinson}, {Rownd}, {Saraceno}, {Sauvage}, {Savage}, {Savini},
  {Sawyer}, {Scharmberg}, {Schmitt}, {Schneider}, {Schulz}, {Schwartz},
  {Shafer}, {Shupe}, {Sibthorpe}, {Sidher}, {Smith}, {Smith}, {Smith},
  {Spencer}, {Stobie}, {Sudiwala}, {Sukhatme}, {Surace}, {Stevens}, {Swinyard},
  {Trichas}, {Tourette}, {Triou}, {Tseng}, {Tucker}, {Turner}, {Vaccari},
  {Valtchanov}, {Vigroux}, {Virique}, {Voellmer}, {Walker}, {Ward}, {Waskett},
  {Weilert}, {Wesson}, {White}, {Whitehouse}, {Wilson}, {Winter}, {Woodcraft},
  {Wright}, {Xu}, {Zavagno}, {Zemcov}, {Zhang}, \& {Zonca}}]{Griffin2010}
{Griffin}, M.~J., {Abergel}, A., {Abreu}, A., {et~al.} 2010, \aap, 518, L3

\bibitem[{{Hindson} {et~al.}(2013){Hindson}, {Thompson}, {Urquhart}, {Faimali},
  {Johnston-Hollitt}, {Clark}, \& {Davies}}]{Hindson2013}
{Hindson}, L., {Thompson}, M.~A., {Urquhart}, J.~S., {et~al.} 2013, \mnras,
  435, 2003

\bibitem[{{Hou} \& {Gao}(2014)}]{Hou2014}
{Hou}, L.~G. \& {Gao}, X.~Y. 2014, \mnras, 438, 426

\bibitem[{{Jackson} {et~al.}(2010){Jackson}, {Finn}, {Chambers}, {Rathborne},
  \& {Simon}}]{Jackson2010}
{Jackson}, J.~M., {Finn}, S.~C., {Chambers}, E.~T., {Rathborne}, J.~M., \&
  {Simon}, R. 2010, \apjl, 719, L185

\bibitem[{{Jones} {et~al.}(1999){Jones}, {Frey}, {Verstraete}, {Cox}, \&
  {Demyk}}]{Jones1999}
{Jones}, A., {Frey}, V., {Verstraete}, L., {Cox}, P., \& {Demyk}, K. 1999, in
  ESA Special Publication, Vol. 427, The Universe as Seen by ISO, ed. P.~{Cox}
  \& M.~{Kessler}, 679

\bibitem[{{Kainulainen} {et~al.}(2013){Kainulainen}, {Ragan}, {Henning}, \&
  {Stutz}}]{Kainulainen2013}
{Kainulainen}, J., {Ragan}, S.~E., {Henning}, T., \& {Stutz}, A. 2013, \aap,
  557, A120

\bibitem[{{Kauffmann} \& {Pillai}(2010)}]{Kauffmann2010}
{Kauffmann}, J. \& {Pillai}, T. 2010, \apjl, 723, L7

\bibitem[{{Kramer} {et~al.}(1998){Kramer}, {Stutzki}, {Rohrig}, \&
  {Corneliussen}}]{Kramer1998}
{Kramer}, C., {Stutzki}, J., {Rohrig}, R., \& {Corneliussen}, U. 1998, \aap,
  329, 249

\bibitem[{{Krumholz} \& {Matzner}(2009)}]{krum2009}
{Krumholz}, M.~R. \& {Matzner}, C.~D. 2009, \apj, 703, 1352

\bibitem[{{Lee} \& {Chen}(2007)}]{Lee2007}
{Lee}, H.-T. \& {Chen}, W.~P. 2007, \apj, 657, 884

\bibitem[{{Lefloch} \& {Lazareff}(1994)}]{Lefloch1994}
{Lefloch}, B. \& {Lazareff}, B. 1994, \aap, 289, 559

\bibitem[{{Li} {et~al.}(2013){Li}, {Wyrowski}, {Menten}, \&
  {Belloche}}]{Li2013}
{Li}, G.-X., {Wyrowski}, F., {Menten}, K., \& {Belloche}, A. 2013, \aap, 559,
  A34

\bibitem[{{Martins} {et~al.}(2010){Martins}, {Pomar{\`e}s}, {Deharveng},
  {Zavagno}, \& {Bouret}}]{Martins2010}
{Martins}, F., {Pomar{\`e}s}, M., {Deharveng}, L., {Zavagno}, A., \& {Bouret},
  J.~C. 2010, \aap, 510, A32

\bibitem[{{Mezger} \& {Henderson}(1967)}]{mezg1967}
{Mezger}, P.~G. \& {Henderson}, A.~P. 1967, \apj, 147, 471

\bibitem[{{Mezger} {et~al.}(1974){Mezger}, {Smith}, \& {Churchwell}}]{mezg1974}
{Mezger}, P.~G., {Smith}, L.~F., \& {Churchwell}, E. 1974, \aap, 32, 269

\bibitem[{{Nagai} {et~al.}(1998){Nagai}, {Inutsuka}, \& {Miyama}}]{Nagai1998}
{Nagai}, T., {Inutsuka}, S.-i., \& {Miyama}, S.~M. 1998, \apj, 506, 306

\bibitem[{{Padoan} {et~al.}(2001){Padoan}, {Juvela}, {Goodman}, \&
  {Nordlund}}]{Padoan2001}
{Padoan}, P., {Juvela}, M., {Goodman}, A.~A., \& {Nordlund}, {\AA}. 2001, \apj,
  553, 227

\bibitem[{{Panagia}(1973)}]{pana1973}
{Panagia}, N. 1973, \aj, 78, 929

\bibitem[{{Peng} {et~al.}(2010){Peng}, {Wyrowski}, {van der Tak}, {Menten}, \&
  {Walmsley}}]{Peng2010}
{Peng}, T.-C., {Wyrowski}, F., {van der Tak}, F.~F.~S., {Menten}, K.~M., \&
  {Walmsley}, C.~M. 2010, \aap, 520, A84

\bibitem[{{Peretto} {et~al.}(2014){Peretto}, {Fuller}, {Andr{\'e}},
  {Arzoumanian}, {Rivilla}, {Bardeau}, {Duarte Puertas}, {Guzman Fernandez},
  {Lenfestey}, {Li}, {Olguin}, {R{\"o}ck}, {de Villiers}, \&
  {Williams}}]{Peretto2014}
{Peretto}, N., {Fuller}, G.~A., {Andr{\'e}}, P., {et~al.} 2014, \aap, 561, A83

\bibitem[{{Poglitsch} {et~al.}(2010){Poglitsch}, {Waelkens}, {Geis},
  {Feuchtgruber}, {Vandenbussche}, {Rodriguez}, {Krause}, {Renotte}, {van
  Hoof}, {Saraceno}, {Cepa}, {Kerschbaum}, {Agn{\`e}se}, {Ali}, {Altieri},
  {Andreani}, {Augueres}, {Balog}, {Barl}, {Bauer}, {Belbachir}, {Benedettini},
  {Billot}, {Boulade}, {Bischof}, {Blommaert}, {Callut}, {Cara}, {Cerulli},
  {Cesarsky}, {Contursi}, {Creten}, {De Meester}, {Doublier}, {Doumayrou},
  {Duband}, {Exter}, {Genzel}, {Gillis}, {Gr{\"o}zinger}, {Henning},
  {Herreros}, {Huygen}, {Inguscio}, {Jakob}, {Jamar}, {Jean}, {de Jong},
  {Katterloher}, {Kiss}, {Klaas}, {Lemke}, {Lutz}, {Madden}, {Marquet},
  {Martignac}, {Mazy}, {Merken}, {Montfort}, {Morbidelli}, {M{\"u}ller},
  {Nielbock}, {Okumura}, {Orfei}, {Ottensamer}, {Pezzuto}, {Popesso},
  {Putzeys}, {Regibo}, {Reveret}, {Royer}, {Sauvage}, {Schreiber}, {Stegmaier},
  {Schmitt}, {Schubert}, {Sturm}, {Thiel}, {Tofani}, {Vavrek}, {Wetzstein},
  {Wieprecht}, \& {Wiezorrek}}]{Poglitsch2010}
{Poglitsch}, A., {Waelkens}, C., {Geis}, N., {et~al.} 2010, \aap, 518, L2

\bibitem[{{Preibisch} {et~al.}(2012){Preibisch}, {Roccatagliata}, {Gaczkowski},
  \& {Ratzka}}]{Preibisch2012}
{Preibisch}, T., {Roccatagliata}, V., {Gaczkowski}, B., \& {Ratzka}, T. 2012,
  \aap, 541, A132

\bibitem[{{Ragan} {et~al.}(2014){Ragan}, {Henning}, {Tackenberg}, {Beuther},
  {Johnston}, {Kainulainen}, \& {Linz}}]{Ragan2014}
{Ragan}, S.~E., {Henning}, T., {Tackenberg}, J., {et~al.} 2014, \aap, 568, A73

\bibitem[{{Rathborne} {et~al.}(2006){Rathborne}, {Jackson}, \&
  {Simon}}]{Rathborne2006}
{Rathborne}, J.~M., {Jackson}, J.~M., \& {Simon}, R. 2006, \apj, 641, 389

\bibitem[{{Simon} {et~al.}(2006){Simon}, {Jackson}, {Rathborne}, \&
  {Chambers}}]{Simon2006}
{Simon}, R., {Jackson}, J.~M., {Rathborne}, J.~M., \& {Chambers}, E.~T. 2006,
  \apj, 639, 227

\bibitem[{{Simpson} {et~al.}(2012){Simpson}, {Povich}, {Kendrew}, {Lintott},
  {Bressert}, {Arvidsson}, {Cyganowski}, {Maddison}, {Schawinski}, {Sherman},
  {Smith}, \& {Wolf-Chase}}]{simp2012}
{Simpson}, R.~J., {Povich}, M.~S., {Kendrew}, S., {et~al.} 2012, \mnras, 424,
  2442

\bibitem[{{Smith} {et~al.}(2014){Smith}, {Glover}, {Clark}, {Klessen}, \&
  {Springel}}]{Smith2014}
{Smith}, R.~J., {Glover}, S.~C.~O., {Clark}, P.~C., {Klessen}, R.~S., \&
  {Springel}, V. 2014, \mnras, 441, 1628

\bibitem[{{Stutzki} \& {Guesten}(1990)}]{Stutzki1990}
{Stutzki}, J. \& {Guesten}, R. 1990, \apj, 356, 513

\bibitem[{{Wang} {et~al.}(2014){Wang}, {Zhang}, {Testi}, {van der Tak}, {Wu},
  {Zhang}, {Pillai}, {Wyrowski}, {Carey}, {Ragan}, \& {Henning}}]{Wang2014}
{Wang}, K., {Zhang}, Q., {Testi}, L., {et~al.} 2014, \mnras, 439, 3275

\bibitem[{{Wang} {et~al.}(2011){Wang}, {Zhang}, {Wu}, \& {Zhang}}]{Wang2011}
{Wang}, K., {Zhang}, Q., {Wu}, Y., \& {Zhang}, H. 2011, \apj, 735, 64

\bibitem[{{Watson} {et~al.}(2008){Watson}, {Povich}, {Churchwell}, {Babler},
  {Chunev}, {Hoare}, {Indebetouw}, {Meade}, {Robitaille}, \&
  {Whitney}}]{wats2008}
{Watson}, C., {Povich}, M.~S., {Churchwell}, E.~B., {et~al.} 2008, \apj, 681,
  1341

\bibitem[{{Weaver} {et~al.}(1977){Weaver}, {McCray}, {Castor}, {Shapiro}, \&
  {Moore}}]{Weaver1977}
{Weaver}, R., {McCray}, R., {Castor}, J., {Shapiro}, P., \& {Moore}, R. 1977,
  \apj, 218, 377

\bibitem[{{Whitworth} {et~al.}(1994{\natexlab{a}}){Whitworth}, {Bhattal},
  {Chapman}, {Disney}, \& {Turner}}]{Whitworth1994}
{Whitworth}, A.~P., {Bhattal}, A.~S., {Chapman}, S.~J., {Disney}, M.~J., \&
  {Turner}, J.~A. 1994{\natexlab{a}}, \aap, 290, 421

\bibitem[{{Whitworth} {et~al.}(1994{\natexlab{b}}){Whitworth}, {Bhattal},
  {Chapman}, {Disney}, \& {Turner}}]{whit1994}
{Whitworth}, A.~P., {Bhattal}, A.~S., {Chapman}, S.~J., {Disney}, M.~J., \&
  {Turner}, J.~A. 1994{\natexlab{b}}, \mnras, 268, 291

\bibitem[{{Winnewisser} {et~al.}(1979){Winnewisser}, {Churchwell}, \&
  {Walmsley}}]{winn1979}
{Winnewisser}, G., {Churchwell}, E., \& {Walmsley}, C.~M. 1979, \aap, 72, 215

\bibitem[{{Yuan} {et~al.}(2014){Yuan}, {Wu}, {Li}, \& {Liu}}]{Yuan2014}
{Yuan}, J.-H., {Wu}, Y., {Li}, J.~Z., \& {Liu}, H. 2014, \apj, 797, 40

\bibitem[{{Zhang} \& {Wang}(2012)}]{s51}
{Zhang}, C.~P. \& {Wang}, J.~J. 2012, \aap, 544, A11

\bibitem[{{Zhang} \& {Wang}(2013)}]{n68}
{Zhang}, C.-P. \& {Wang}, J.-J. 2013, Research in Astronomy and Astrophysics,
  13, 47

\bibitem[{{Zhang} {et~al.}(2013){Zhang}, {Wang}, \& {Xu}}]{n131}
{Zhang}, C.-P., {Wang}, J.-J., \& {Xu}, J.-L. 2013, \aap, 550, A117

\end{thebibliography}

\end{document}